\documentclass[]{aiaa-tc}

\usepackage{subfigure}
\usepackage{floatrow}
\usepackage{wrapfig}
\usepackage{amsfonts,amsmath,amsthm,mathrsfs}
\usepackage{amssymb}		
\usepackage{array}
\usepackage{multicol}
\usepackage{url}

\newcommand{\degrees}{\ensuremath{^\circ}}
\newfloatcommand{capbtabbox}{table}[][\FBwidth]

 \title{Finite Element Flow Simulations of the EUROLIFT DLR-F11 High Lift Configuration}

 \author{
  Kedar C. Chitale\thanks{Postdoctoral Research Associate, Rensselaer Polytechnic Institute, AIAA Member, Tel.: +1-518-596-4750} \\ 
  {\normalsize\itshape
   MANE Dept., Rensselaer Polytechnic Institute, NY 12180} \\
   \and
     Michel Rasquin\thanks{Postdoctoral Research Associate, Argonne National Laboratory} \\
      {\normalsize\itshape Leadership Computing Facility, Argonne National Laboratory, IL 60439}
  \and
       Jeffrey Martin \\
      {\normalsize\itshape Massachusetts Institute of Technology, MA 02139}
  \and
   Kenneth E. Jansen\thanks{Professor, University of Colorado Boulder, AIAA Associate Fellow, Tel.: +1-303-492-4359}\\
   {\normalsize\itshape
   Dept. of Aerospace Engineering Sciences, University of Colorado Boulder, Boulder, CO, 80309-0429}\\
 }

 \AIAApapernumber{YEAR-NUMBER}
 \AIAAconference{Conference Name, Date, and Location}
 \AIAAcopyright{\AIAAcopyrightD{YEAR}}

\begin{document}

\maketitle

\begin{abstract}

This paper presents flow simulation results of the EUROLIFT DLR-F11 multi-element wing configuration, obtained with a highly scalable finite element solver, PHASTA. This work was accomplished as a part of the $2^{\text{nd}}$ high lift prediction workshop. In-house meshes were constructed with increasing mesh density for analysis. A solution adaptive approach was used as an alternative and its effectiveness was studied by comparing its results with the ones obtained with other meshes. Comparisons between the numerical solution obtained with unsteady RANS turbulence model and available experimental results are provided for verification and discussion. Based on the observations, future direction for adaptive research and simulations with higher fidelity turbulence models is outlined. 

\end{abstract}

\section*{Nomenclature}

\begin{tabbing}
  XXXXX \= \kill
  $ y^+$ \> Dimensionless distance from the wall ($ \rho u_\tau y/\mu$) \\
  $ Re$ \> Reynolds number \\
  $ C_p$ \> Coefficient of pressure \\
  $C_L$ \> Coefficient of lift \\
  $ C_D$ \> Coefficient of drag \\
  $ \alpha$ \> Angle of attack (AoA)\\
  $ \Delta t$ \> Time step size ($s$) \\
  $ b$ \> Span of the wing $m$
  
 \end{tabbing}

\section{Introduction}

Multi-element wing systems are popular among aerospace communities because of their ability to generate increased lift.  Most modern aircrafts use some version of these systems and have flaps and slats to boost the lifting performance during takeoff and landing. However, these systems present a challenging problem for high Reynolds number simulations. 

Over the years, unstructured meshes have become increasingly popular in the CFD community due to their ease of adjusting to complex flow domains. Multi-element wings are one class of such geometries which present a challenging scenario for mesh generation. Moreover, since high Reynolds numbers are considered for these simulations ($Re = 1.35$M and $Re = 15.1$M), meshes with tight mesh spacings near the walls need to be used, in order to satisfy the requirements posed by the turbulence models. To achieve these fine mesh spacings, layered structured mesh is used near the no-slip walls, in the boundary layer region. The meshing guidelines for the $2^{\text{nd}}$ high lift prediction workshop~\cite{HiLiftPW2} also suggest using semi-structured mesh inside the boundary layer region.

The required resolution in areas of interest like wakes and tips of wings is usually not known $a priori$ to the simulations. As a result, a way to achieve an adequate level of resolution in these regions of the domain is to progressively refine the initial mesh. However, such an approach is not ideal and an alternative efficient automatic adaptive approach is more suitable, which would refine only in the areas pointed out by a suitable error indicator/estimator. When using unstructured meshes with layered mesh near the walls, adaptivity becomes more complex. In addition to using progressively refined meshes, we also use an adaptive approach suitable for such meshes for comparison. 

The turbulence model used in this work is RANS Spalart-Allmaras (SA)\cite{SA}. It was shown in the $ 1^{\text{st}}$ high lift prediction workshop that the results obtained with Spalart-Allmaras model were in good agreement with the experiments\cite{HLPW1}. No wall models were used which leads to a wall resolved approach requiring the use of very tight mesh spacings for the first layer off the wall. Generally, a distance of $ y^+=1$ for the first layer is considered enough in the case of SA turbulence model with a wall resolved treatment. 

\begin{figure}[h!]
\begin{center}
\subfigure[Configuration 2]{
	\fbox{\includegraphics[width=7.5cm]{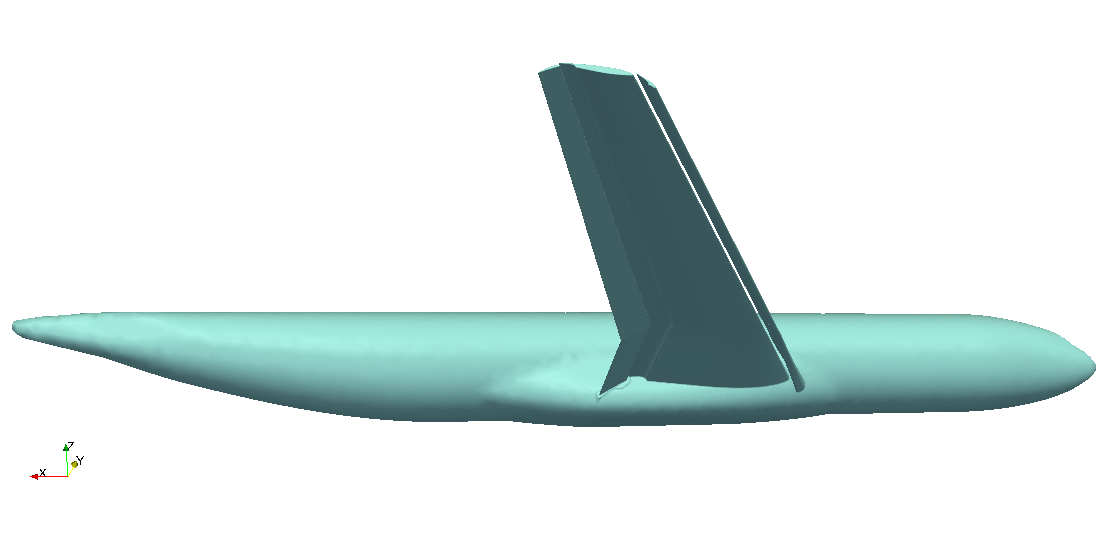}}
	\label{f:Config2Geom}
}
\subfigure[Configuration 4]{
	\fbox{\includegraphics[width=7.5cm]{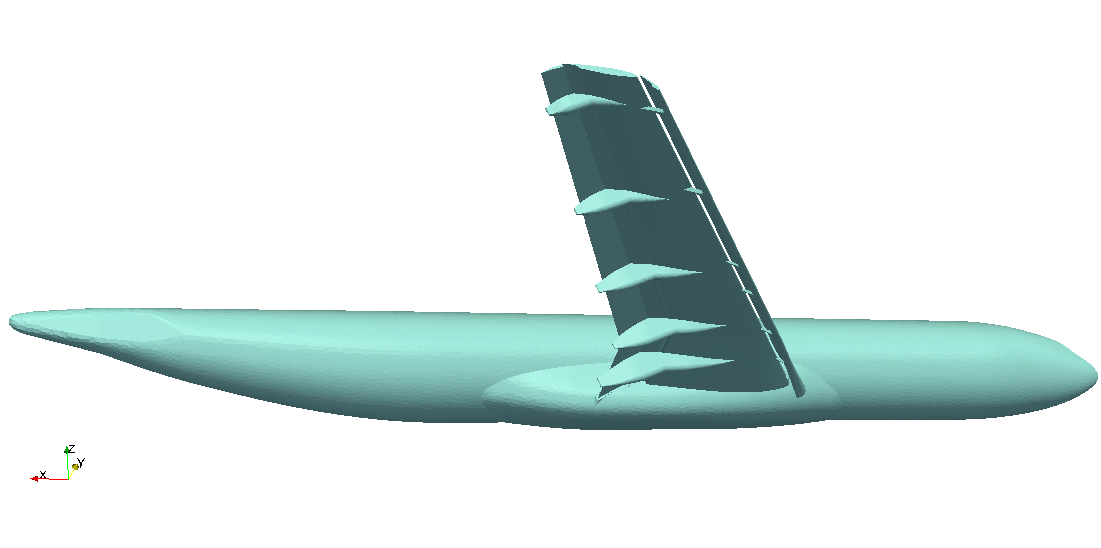}}
	\label{f:Config4Geom}
}
\vspace{-10pt}
\caption{Different configurations of DLR-F11 used for analysis}
\label{f:Geometries}
\end{center}
\end{figure}

The simulations were carried out for 2 different studies. First, a grid convergence study was considered (Case 1), whereas the second one consisted of a Reynolds number study (Case 2). The geometries for both these studies are displayed in Figure~\ref{f:Geometries}. The Configuration 2 geometry consists of a main wing attached to a long realistic fuselage along with slats and flaps in landing configuration. The nominal slat angle is 26.5\degrees \hspace{0.1mm} and the nominal flap angle is 32\degrees. In addition to these elements, Configuration 4 geometry consists of slat tracks and flap track fairings. Additional geometry details are given in Table~\ref{t:DLRGeometry} and in Rudnick et al.\cite{HiLiftRudnick}.

The article is organized as follows. Section~\ref{s:meshgen} describes the characteristics of the meshes generated for this work. The mesh adaption strategy is then briefly presented in Section~\ref{s:adaptiveanalysis}, whereas the details of our finite element flow solver are given in Section~\ref{s:flowsolver}. The sensitivity of the results with respect to the grid resolution and the influence of the Reynolds number are both discussed in Section~\ref{s:results}. Finally, the conclusions are presented in Section~\ref{s:conclusions}.

\begin{table}[h!]
\centering
\newcolumntype{A}{>{\centering\arraybackslash}m{3 cm}}
\newcolumntype{B}{>{\centering\arraybackslash}m{4 cm}}
\newcolumntype{D}{>{\centering\arraybackslash}m{2 cm}}
      \begin{tabular}{|A|A|A|A|}
      \hline
  	 Mach number & Mean aerodynamic chord (MAC) & semi span (b/2) & Wing aspect ratio \\ \hline
	0.175 &  0.347 $ m$ & 1.4 $m$ & 9.353  \\ \hline
      \end{tabular}
  \caption{Geometric details of DLR-F11 high lift configuration}
  \label{t:DLRGeometry}
\end{table}



\section{Mesh generation}
\label{s:meshgen}

As per the meshing guidelines and due to the specific requirements of the flow solver, three in-house grids were constructed for grid convergence study with Config. 2 (Case 1), coarse, medium and fine. For Reynolds number study (Case 2) with Config. 4, a mesh similar to the medium mesh of Config. 2 in terms of grid spacings was constructed. All of the meshes were constructed using commercial meshing package provided by Simmetrix Inc.\cite{Simmetrix}. The details of various meshes are given in Table~\ref{t:DLRF11Meshes}.

\begin{table}[h!]
\caption[Details and comparison of meshes for the Case 1]{Details of the meshes for Case 1}
\newcolumntype{A}{>{\centering\arraybackslash}m{4 cm}}
\newcolumntype{B}{>{\centering\arraybackslash}m{3 cm}}
\begin{tabular}{|A|B|B|B|}
\hline
 Mesh & \# of elements (million) & \# of vertices  (million) & First cell height (m)\\
\hline
Coarse mesh& 32.28 & 13.55 & 5.5e-7  \\
\hline
Medium mesh & 91.55 & 37.34 & 3.7e-7 \\
\hline
Fine mesh & 287.89 & 112.96 & 2.4e-7 \\
\hline
\end{tabular}
\label{t:DLRF11Meshes}
\end{table}

\begin{figure}[h!]
\begin{center}
\hspace{-10pt}
\subfigure{
	\fbox{\includegraphics[width=5.2cm]{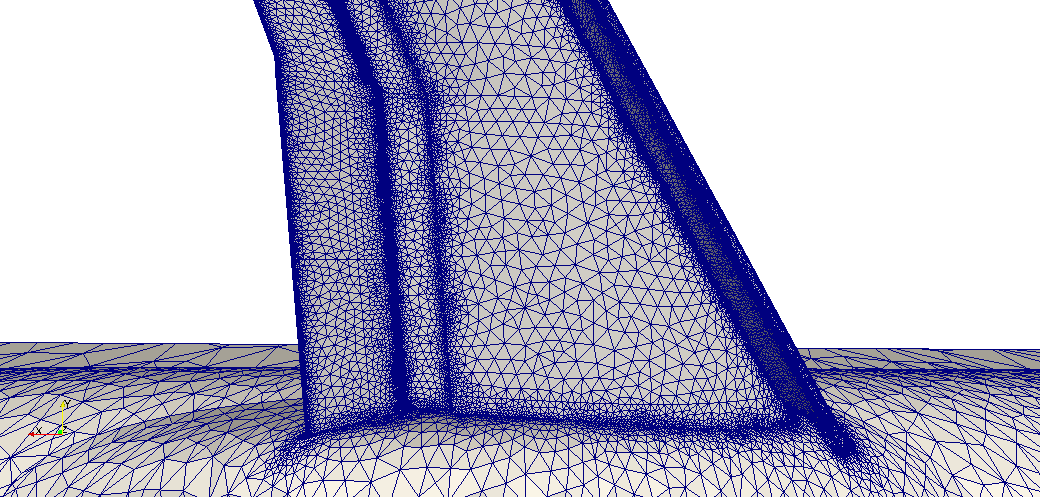}}
}
\subfigure{
	\fbox{\includegraphics[width=5.2cm]{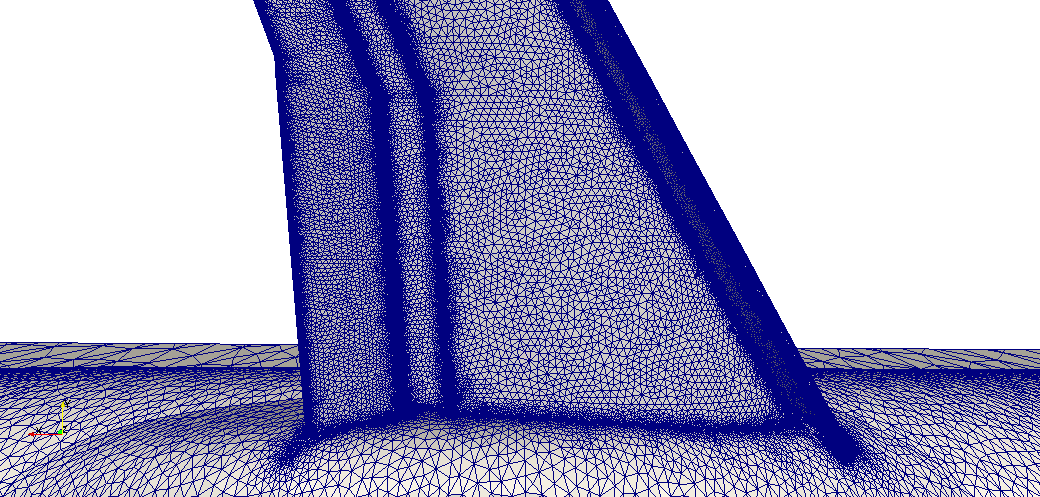}}
}	
\subfigure{
	\fbox{\includegraphics[width=5.2cm]{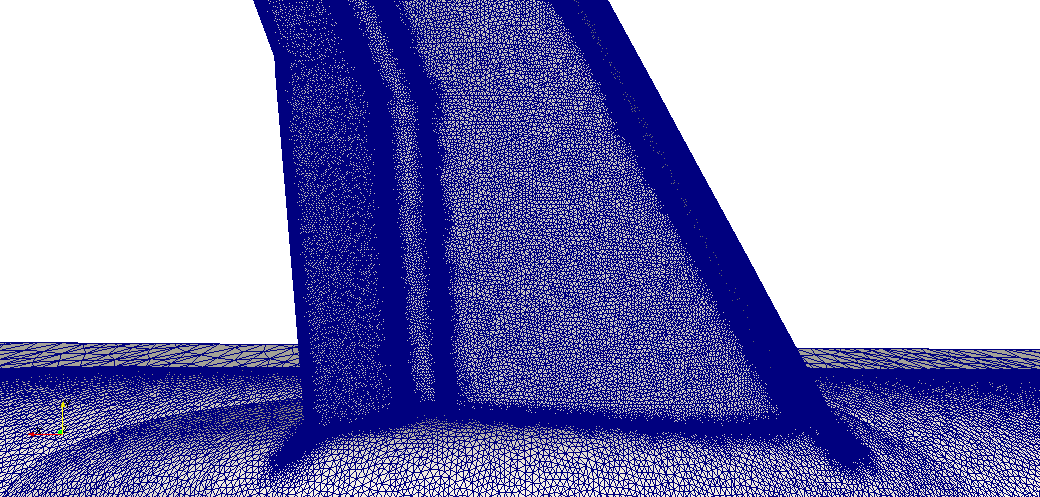}}
}	

%
\subfigure{
	\fbox{\includegraphics[width=5.1cm]{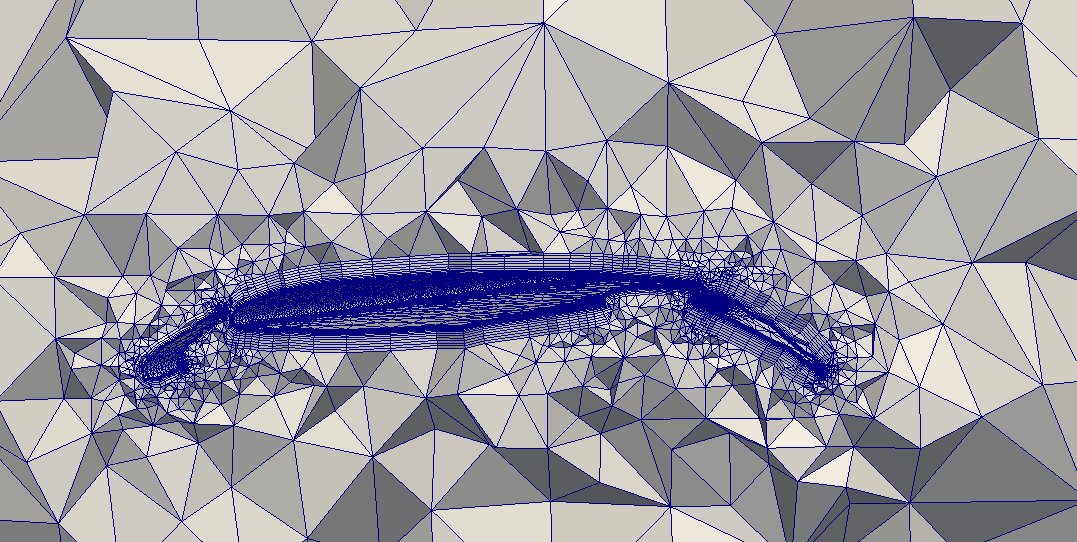}}
}
\subfigure{
	\fbox{\includegraphics[width=5.1cm]{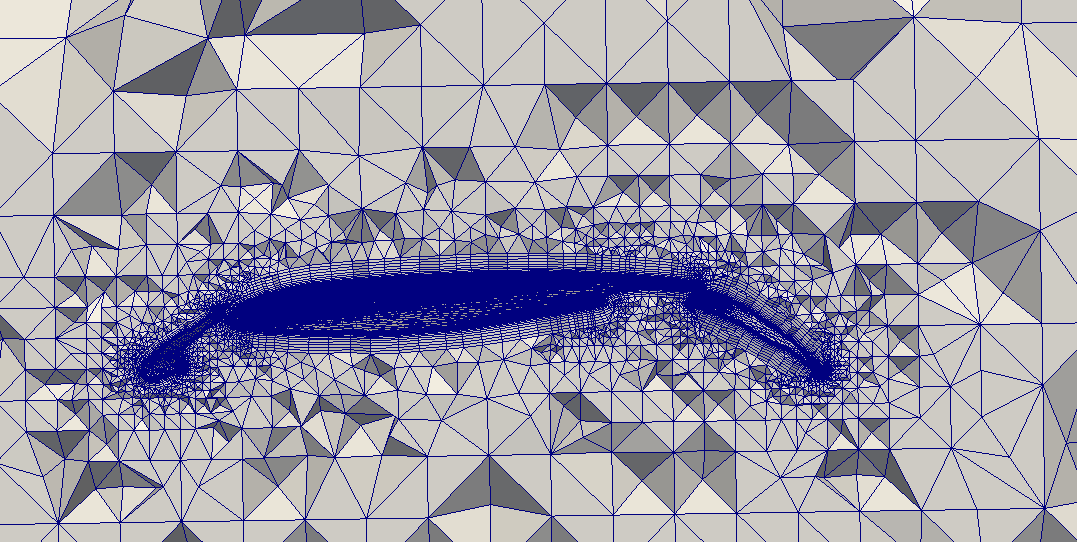}}
}	
\subfigure{
	\fbox{\includegraphics[width=5.4cm]{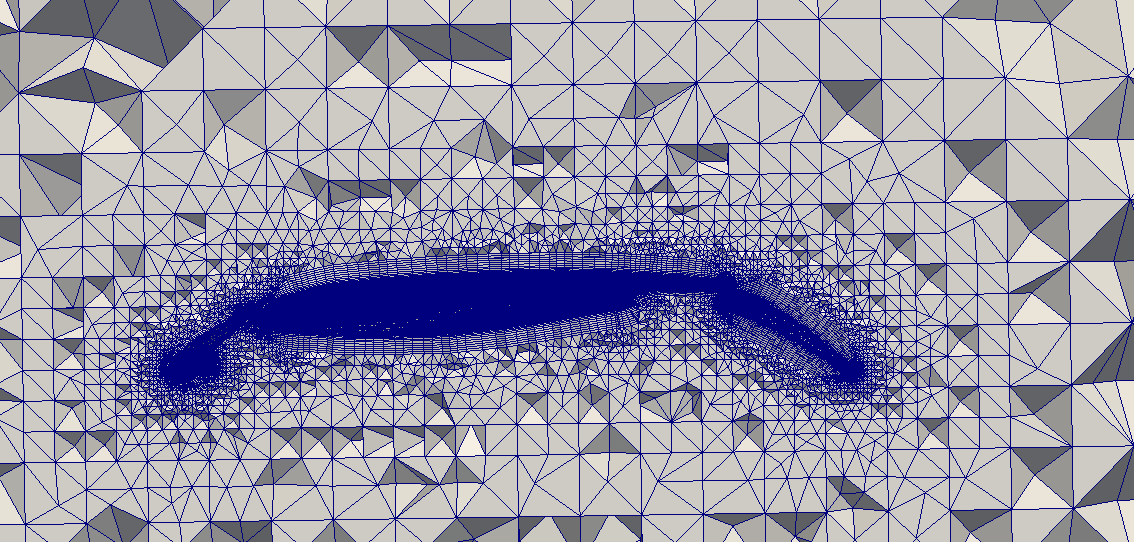}}
}	

\subfigure{
	\fbox{\includegraphics[width=5.1cm]{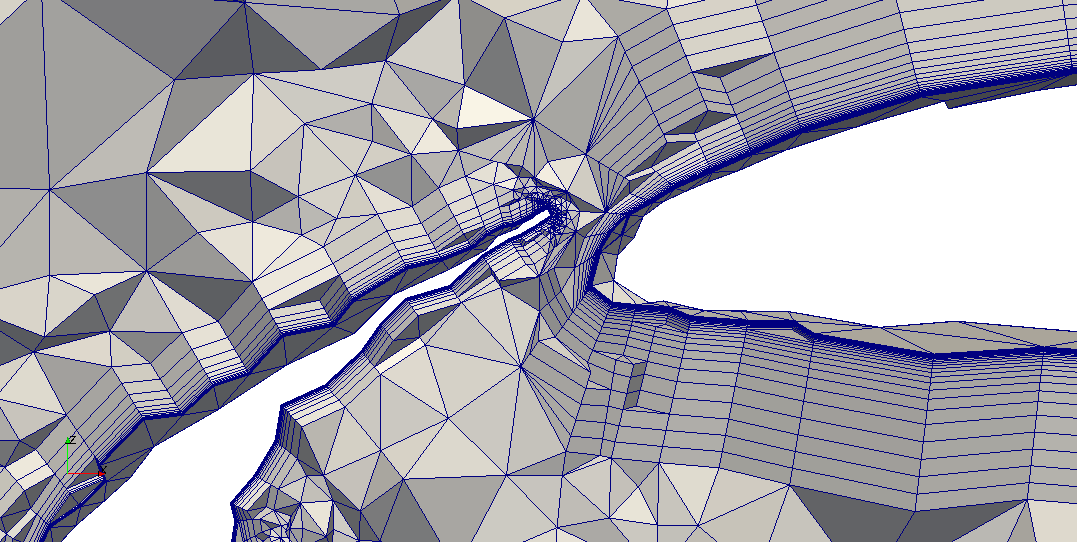}}
}
\subfigure{
	\fbox{\includegraphics[width=5.1cm]{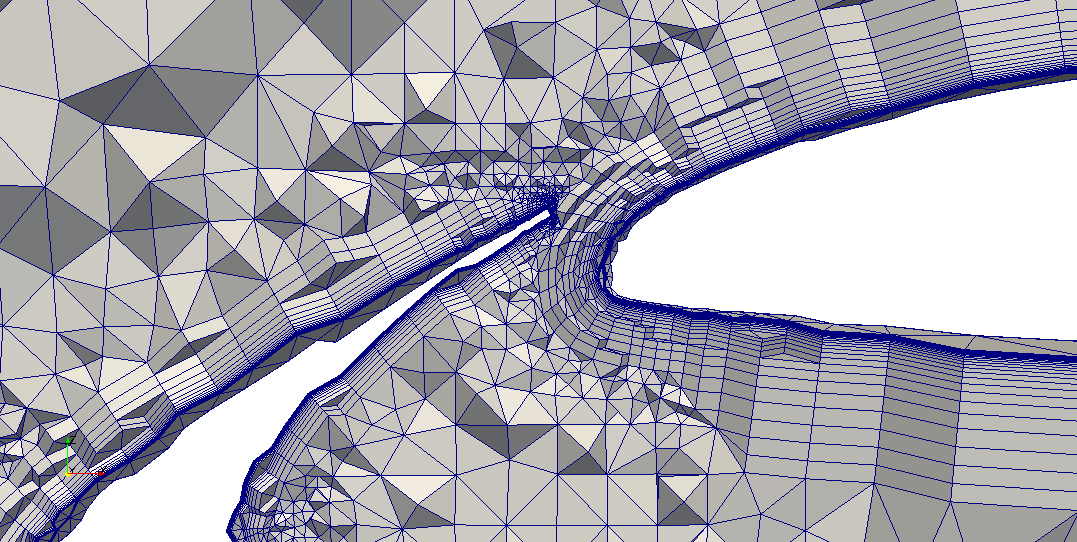}}
}	
\subfigure{
	\fbox{\includegraphics[width=5.4cm]{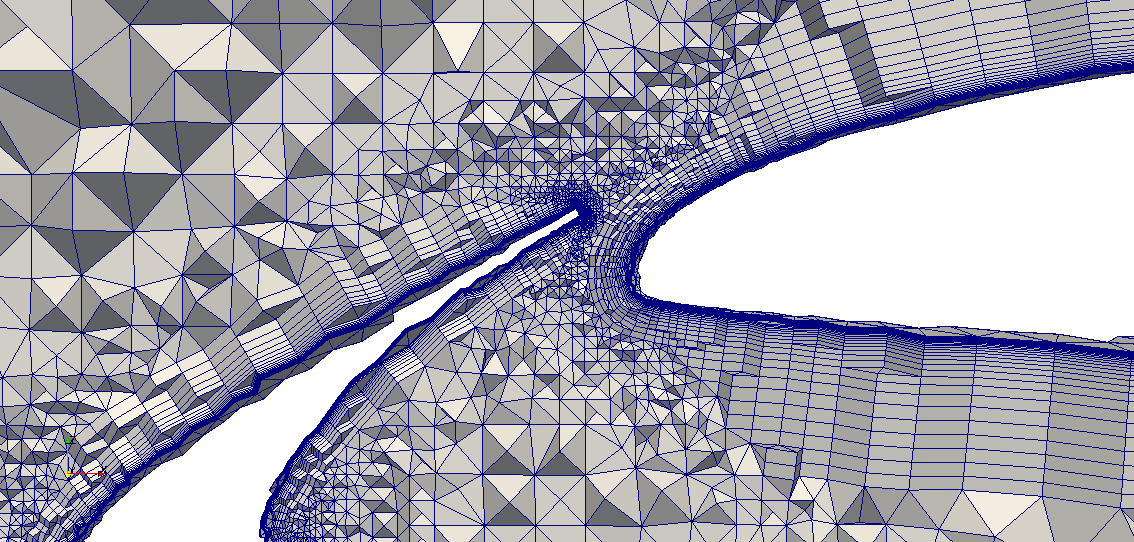}}
}	

\hspace{-10pt}
\subfigure[Coarse]{
	\fbox{\includegraphics[width=5.1cm]{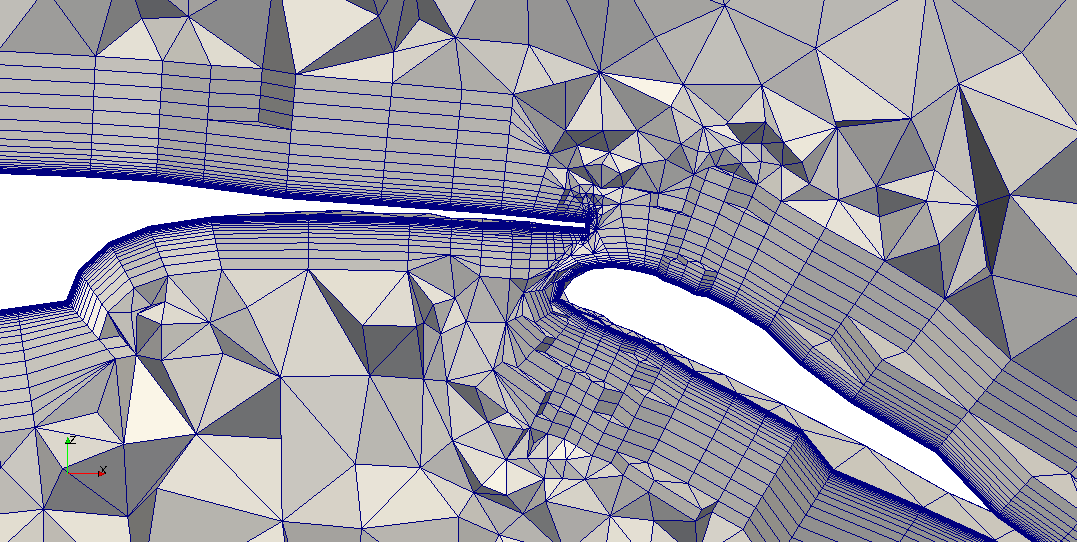}}
	\label{f:DLRConfig2}
}
\hspace{-5pt}
\subfigure[Medium]{
	\fbox{\includegraphics[width=5.1cm]{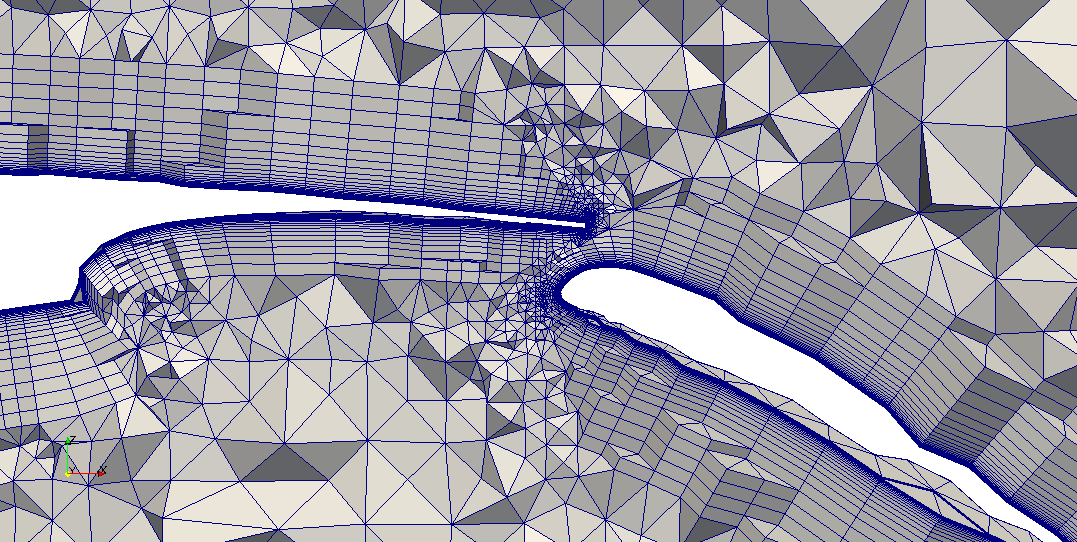}}
	\label{f:DLRConfig4}
}	
\hspace{-5pt}
\subfigure[Fine]{
	\fbox{\includegraphics[width=5.35cm]{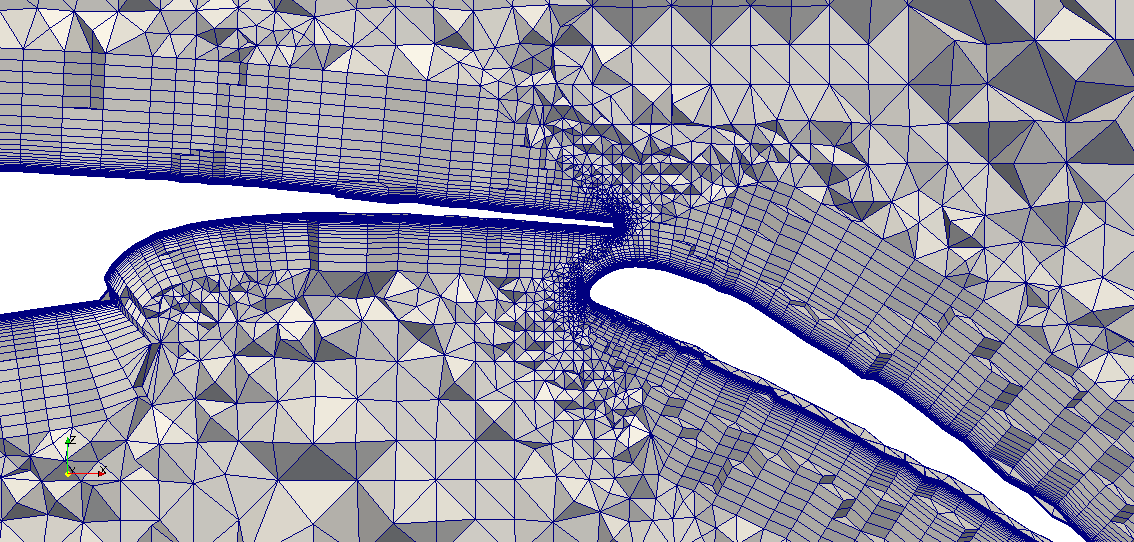}}
	\label{f:DLRConfig4}
}	
\end{center}
 \vspace{-15pt}
 \caption{Mesh pictures of the different meshes for Case 1; first row: main wing, second row: Cut view at mid span, third row: Zoom near nose, fourth row: zoom near trailing edge}
\label{f:DLRMeshes}
\end{figure}

Figure~\ref{f:DLRMeshes} shows several pictures of the meshes for comparison. The first row gives an idea of the surface mesh on the wing and increasing refinement. The second row shows cut view pictures of the meshes at mid-span section to illustrate the increasing mesh density in successive meshes. The third row shows cut views of the meshes near the nose of the main wing and the trailing edge of the slat. Increasing curvature refinement and increasing resolution in the trailing edge region can be clearly seen from these pictures. The fourth row shows cut views of the meshes near the trailing edge of the main wing and the nose of the flap to display progressive refinement in the boundary layer region. 

\begin{figure}[h!]
\begin{center}
\subfigure[View of the wing mesh for Case 2] {
 \fbox{\includegraphics[width = 7.5 cm]{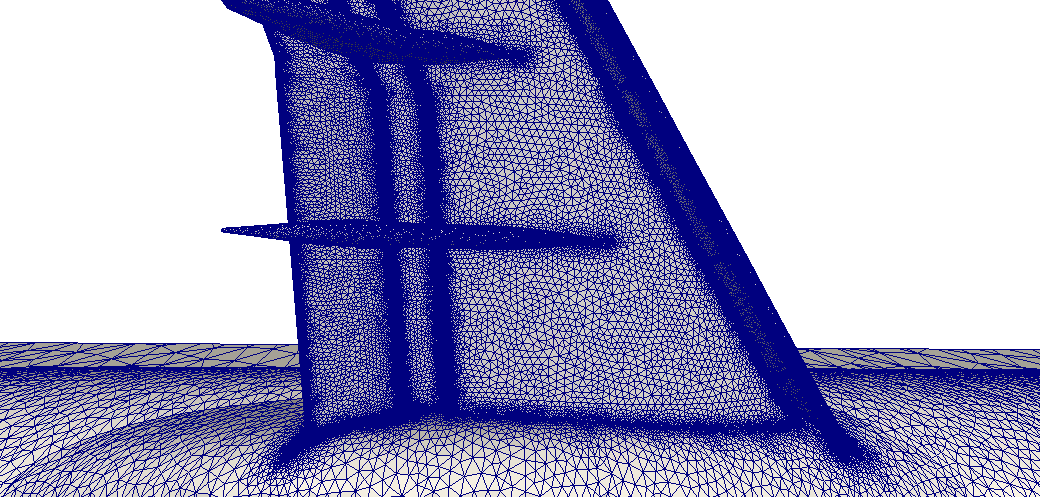}}
 \label{f:Mesh2a_mesh}
}
\subfigure[Zoom of a flap fairing's mesh] {
 \fbox{\includegraphics[width = 7.5 cm]{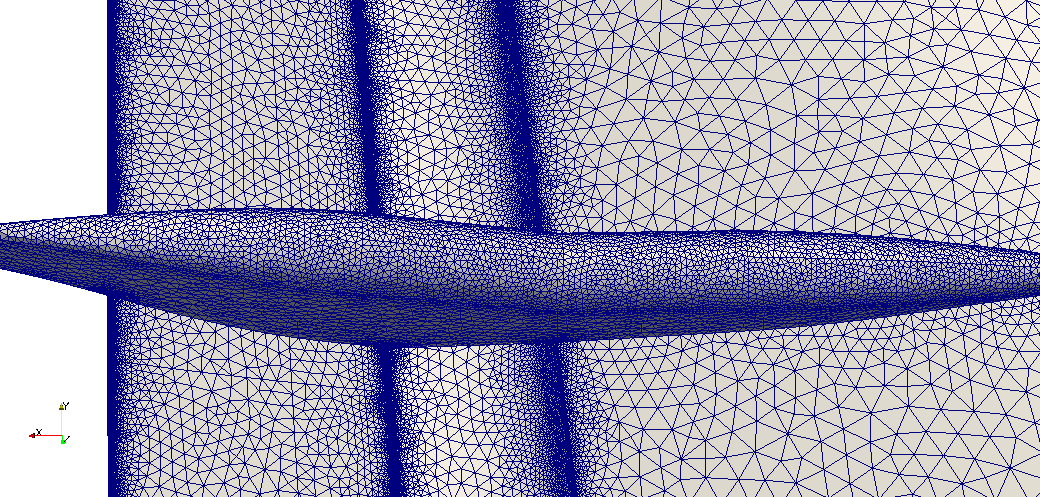}}
 \label{f:Mesh2a_flap}
 }
 \vspace{-15pt}
 \caption{Mesh pictures for Case 2}
 \label{f:Mesh2a}
 \end{center}
\end{figure}

Figure~\ref{f:Mesh2a_mesh} shows the mesh on the pressure side of the wing. The mesh sizes (grid spacings) are similar to the medium mesh of Case 1. Figure~\ref{f:Mesh2a_flap} shows zoom of the mesh on one of the flap fairings. To capture the curvature, the mesh sizes on flap fairings were kept smaller than mesh sizes on the wing. Table~\ref{t:DLRF11Mesh2a} shows the details of the mesh for Case 2. 

\begin{table}[h!]
\caption[Details of the mesh for Case 2]{Details of the mesh for Case 2}
\newcolumntype{A}{>{\centering\arraybackslash}m{4 cm}}
\newcolumntype{B}{>{\centering\arraybackslash}m{3 cm}}
\begin{tabular}{|A|B|B|B|}
\hline
 Mesh & \# of elements (million) & \# of vertices  (million) & First cell height (m)\\
\hline
Medium mesh & 97.90 & 39.88 & 3.7e-7 \\
\hline
\end{tabular}
\label{t:DLRF11Mesh2a}
\end{table}
 
\section{Adaptive analysis}
\label{s:adaptiveanalysis}

Mesh adaptation for such geometries is highly desirable as it has the potential to reduce the computational resources required to efficiently simulate the flow. However, due to the complex nature of such geometries, boundary layer mesh adaptation encounters severe difficulties due to the requirement to adjust to the domain boundary surfaces. When using boundary layer meshes as used in this analysis, it is required to either maintain the layered structure during adaptivity or adapt the wall normal direction in such a way so as to create acceptable meshes for the turbulence model. Previous attempts have been made to use flow physics to drive boundary layer mesh adaptation~\cite{ChitaleBL} for simpler meshes but such an approach is under development and testing for complex geometries like multi-element wings.
 
Boundary layer mesh adaptivity based on Hessians has previously been carried out for viscous flows maintaining the layered structure~\cite{Sahni}. This approach was recently extended to distributed systems~\cite{Ovcharenko}. Previous analysis~\cite{ChitaleMEW} has shown that adaptivity provides an efficient way to reach the required mesh resolution instead of progressive refinement of the initial mesh. 

In this work we use Hessian based anisotropic adaptive techniques applied to an extra coarse mesh in order to create a different set of meshes for the first study and compare the results with that of the other meshes. A combination of total pressure Hessians and solution residual are used to drive adaptivity. Solution residuals drive the smallest mesh sizes and the Hessian provides the directions and relative scales for anisotropic adaptation. The comparison of the adapted meshes is given in Table~\ref{t:DLRF11AdaptMeshes}. The total number of elements in these meshes are on the order of the coarse mesh used for Case 1.

\begin{table}[h!]
\caption[Adapted mesh for DLR-F11]{Adapted meshes for Case 1}
\newcolumntype{A}{>{\centering\arraybackslash}m{4 cm}}
\newcolumntype{B}{>{\centering\arraybackslash}m{3 cm}}
\begin{tabular}{|A|B|B|B|}
\hline
 Mesh & \# of elements (million) & \# of vertices  (million) & First cell height (m)\\
\hline
Adapted: 7Deg & 40.69 & 14.11 & 5.5e-7 \\
\hline
Adapted: 16Deg & 35.0 & 12.04 & 5.5e-7 \\
\hline
\end{tabular}
\label{t:DLRF11AdaptMeshes}
\end{table}

\begin{figure}[h!]
\begin{center}
\subfigure[View of the wing mesh] {
 \fbox{\includegraphics[width =7 cm]{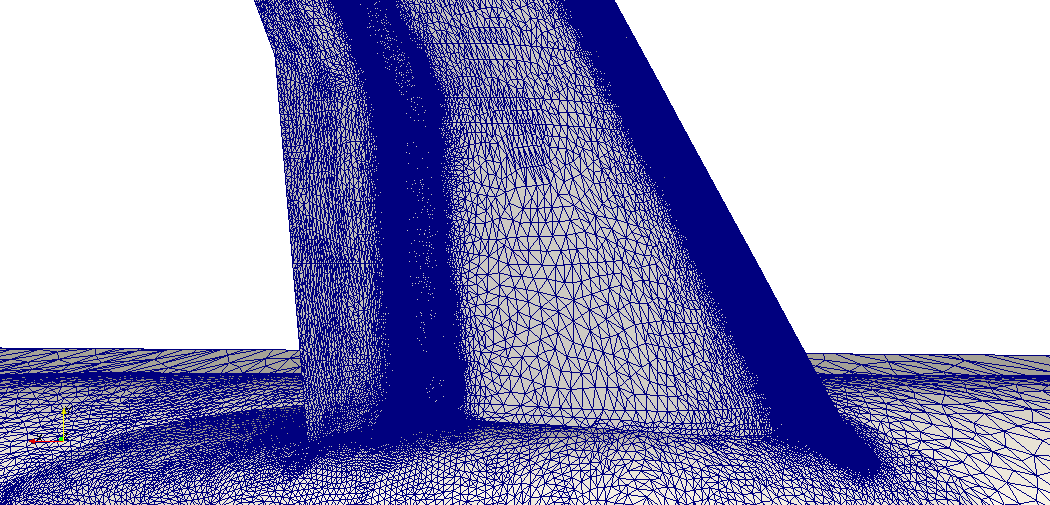}}
 \label{f:MeshAdapt_1}
}
\subfigure[Cut view at mid span] {
 \fbox{\includegraphics[width = 7 cm]{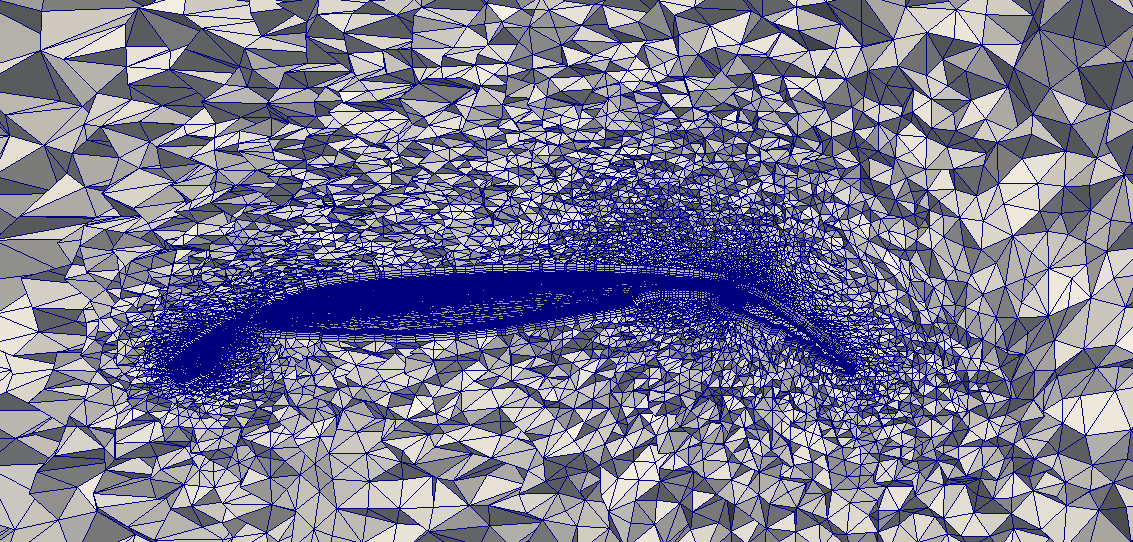}}
 \label{f:MeshAdapt_2}
 }
 \subfigure[Zoom of main wing's nose] {
 \fbox{\includegraphics[width = 7 cm]{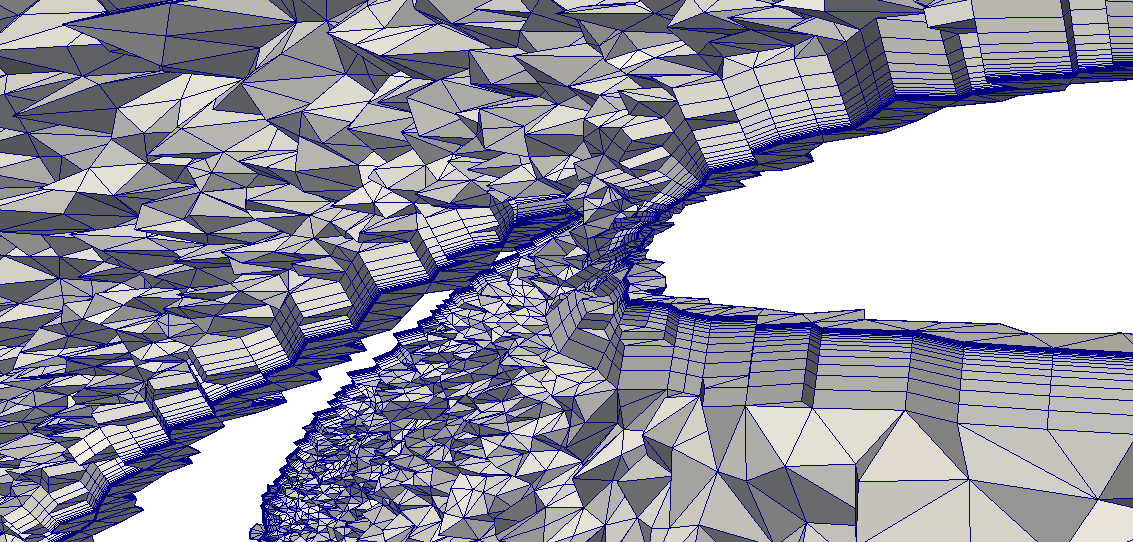}}
 \label{f:MeshAdapt_3}
 }
 \subfigure[Zoom of main wing's trailing edge ] {
 \fbox{\includegraphics[width = 7 cm]{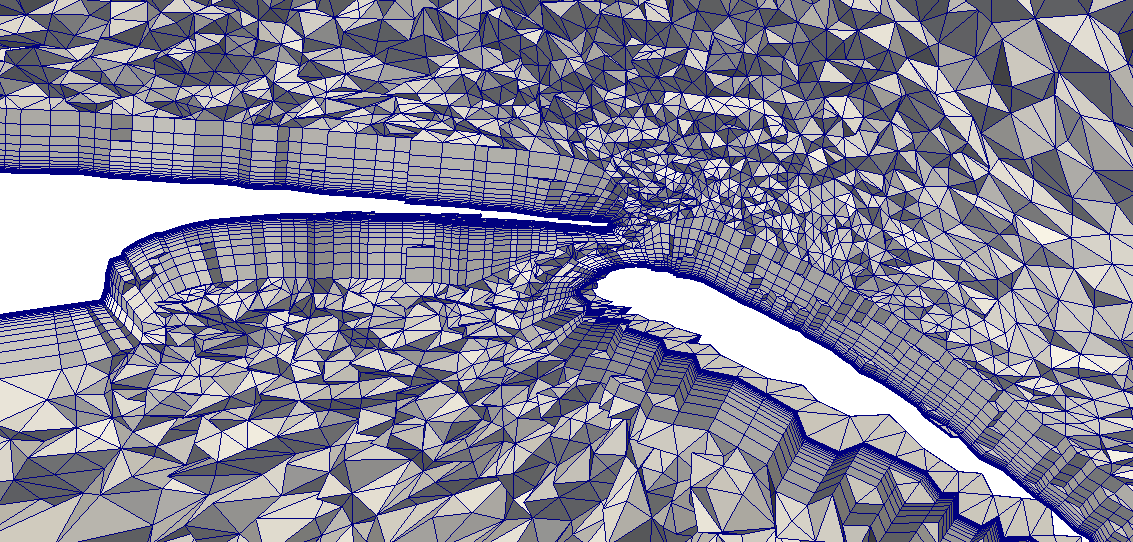}}
 \label{f:MeshAdapt_4}
 }
 \vspace{-20pt}
 \caption{Mesh pictures of the adapted mesh for 7\degrees \hspace{0.1mm} AoA from Case 1}
 \label{f:MeshAdapt7Deg}
 \end{center}
\end{figure}

Figure~\ref{f:MeshAdapt7Deg} shows some mesh pictures of the adapted mesh for 7\degrees \hspace{0.1mm} AoA. The anisotropy in the elements can be seen clearly in all the pictures. On the wing, the elements are stretched in the spanwise direction and the cut view shows elements stretched in the streamwise directions. In general adaptivity shows better resolution in the wake of the wing. 

\section{Details of the flow solver}
\label{s:flowsolver}

The flow solver used in this work is based on a massively parallel stabilized finite element formulation of the incompressible Navier-Stokes equations~\cite{WhiJan99}.  In particular, we employ the streamline upwind/Petrov-Galerkin (SUPG) stabilization method introduced in~\cite{BroHug82} to discretize the governing equations. The stabilized finite element formulation utilized in this work has been shown to be robust, accurate and stable on a variety of flow problems (see for example~\cite{TayHugZar98,WhiJan99}).
In our flow solver (PHASTA), the Navier-Stokes equations (conservation of mass, momentum and energy) plus any auxiliary equations (as needed for turbulence models or level sets in two-phase flow) are discretized in space and time.
The discretization in space based on a stabilized finite element method leads to a weak form of the governing equations, where the solutions (and weight functions) are first interpolated using hierarchic, piecewise polynomials~\cite{WhiJan99,WhiJanDey99}, and followed by the
computation of integrals appearing in the weak form using Gauss quadrature. However, $C^0$ piecewise linear finite elements were considered for this analysis. 
The time integration is based on an implicit generalized-$\alpha$ method~\cite{JanWhi99} which is second order accurate and the resulting algebraic system of equations is solved using Krylov iterative procedures. 
The solver supports various turbulence models like RANS, LES, DES. The flow simulations in this work were performed with unsteady the RANS-SA~\cite{SA} turbulence model.
The simulations were carried out on massively parallel systems. The solver has been shown to scale up to 160K cores~\cite{SahniScaling} and efforts are currently underway to test its performance on 3M MPI processes. The simulations were performed on Janus supercomputer at University of Colorado Boulder and Mira supercomputer (BG/Q) at Argonne National Laboratory. The total number of cores and average number of elements per core are given in Table~\ref{t:MeshRun} for various meshes. 

\begin{table}[h!]
\caption[]{Computational details of Case 1: 7\degrees \hspace{0.1mm} AoA}
\newcolumntype{A}{>{\centering\arraybackslash}m{4 cm}}
\newcolumntype{B}{>{\centering\arraybackslash}m{3 cm}}
\newcolumntype{C}{>{\centering\arraybackslash}m{2 cm}}
\begin{tabular}{|C|B|C|B|}
\hline
 Mesh & \# of elements (million) & \# cores & Average \# elements/core \\ 
\hline
Coarse & 32.28 & 1.8k & 17930\\
\hline
Medium &  91.55 & 32k & 2790 \\
\hline
Fine & 287.89 & 64k &  4390 \\
\hline
Adapted & 40.69 & 3.6k & 10100\\
\hline
\end{tabular}
\label{t:MeshRun}
\end{table}


\section{Results}
\label{s:results}

\subsection{Grid convergence study (Case 1)}

The grid convergence study presented in this section consists of simulating the flow on increasingly refined meshes for a two angles of attack: 7\degrees \hspace{0.1mm}  and 16\degrees. The Reynolds number was kept constant at 15.1 million. Configuration 2 of the wing without any flap track fairings and slat tracks was used in this study. Adapted meshes were also constructed from an initial extra coarse mesh using error indicators extracted from the solution. 

\begin{figure}[h!]
\begin{center}
	\includegraphics[width=12cm]{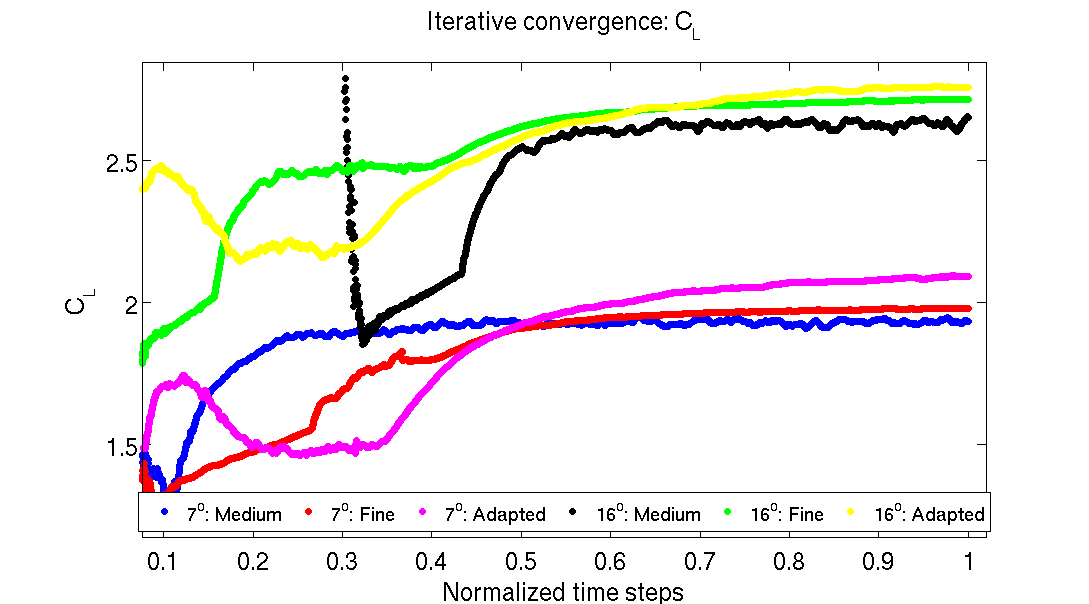}
	\vspace{-10pt}
	\caption{Convergence of $ C_L$ for Case 1}
	\label{f:Config2_IterConvCL}
\end{center}
\end{figure}

Figure~\ref{f:Config2_IterConvCL} shows the iterative convergence of Case 1 simulations for the medium and the fine mesh. The x-axis is normalized by the total number of time steps. The simulations were run till $ C_L$ converged to a stable value and a reasonable time average solution was obtained. The plots show the curves leveling off, indicating that mean quantities can be extracted for $ C_L$. 


\subsubsection{Lift and drag coefficients}

\begin{figure}[h!]
\begin{center}
\subfigure[$ \mathrm{C_L \, vs. \: N^{-2/3}}$]{
	\includegraphics[width=8cm]{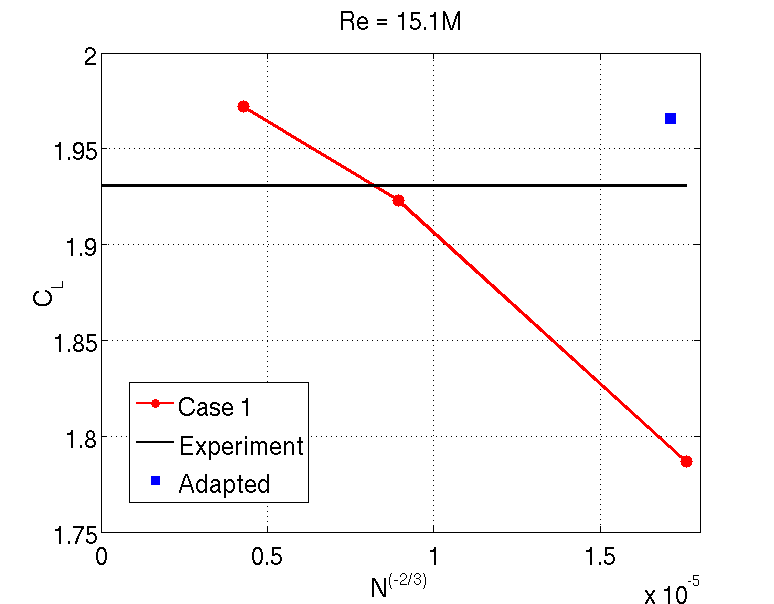}
	\label{f:CLGC7Deg}
}
\hspace{-25pt}
\subfigure[$ \mathrm{C_D \, vs. \: N^{-2/3}}$]{
	\includegraphics[width=8cm]{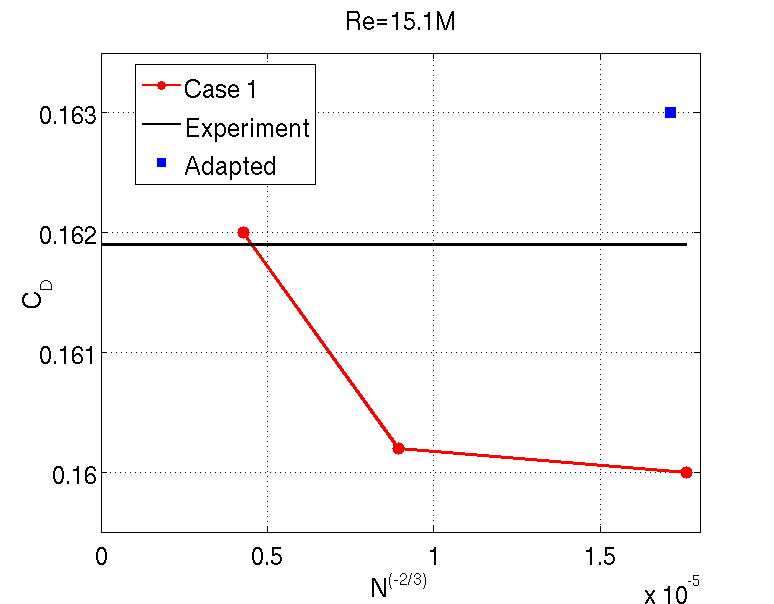}
	\label{f:CDGC7Deg}
}
\end{center}
\vspace{-15pt}
\caption{Grid convergence for 7\degrees \hspace{0.1mm} AoA (Case 1)}
\label{f:GridConv7Deg}
\end{figure}

\begin{figure}[h!]
\begin{center}
\subfigure[$ \mathrm{C_L \, vs. \: N^{-2/3}}$]{
	\includegraphics[width=8cm]{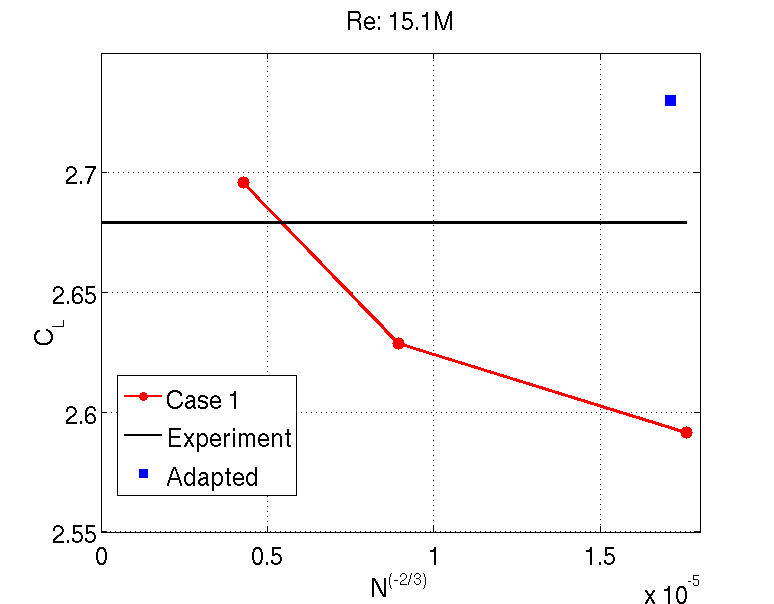}
	\label{f:CLGC16Deg}
}
\hspace{-25pt}
\subfigure[$ \mathrm{C_D \, vs. \: N^{-2/3}}$]{
	\includegraphics[width=8cm]{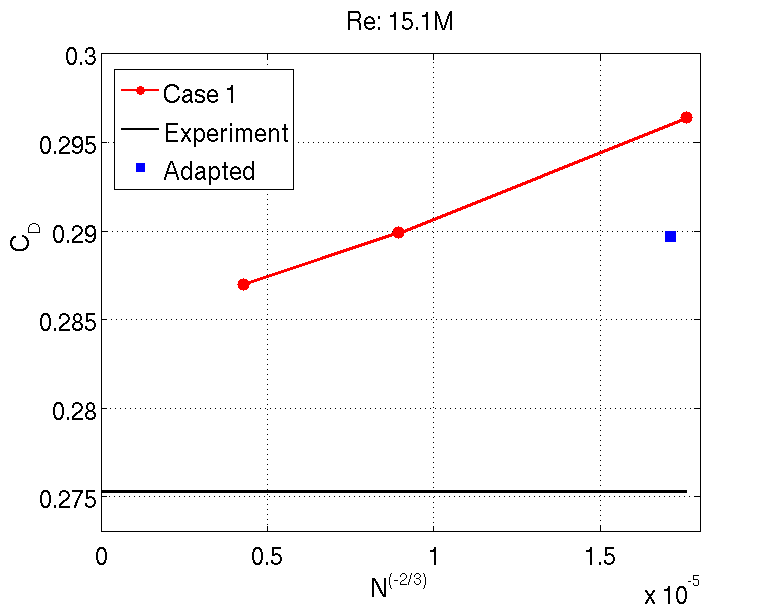}
	\label{f:CDGC16Deg}
}
\end{center}
\vspace{-15pt}
\caption{Grid convergence for 16\degrees \hspace{0.1mm} AoA (Case 1)}
\label{f:GridConv16Deg}
\end{figure}

The purpose of this study is to establish grid convergence as one moves from coarser to finer meshes. Figures~\ref{f:GridConv7Deg} and~\ref{f:GridConv16Deg} plots $ C_L$ and $ C_D$ as a function of $N^{-2/3}$ for 7\degrees \hspace{0.1mm} and 16\degrees, where N is the number of node points. As one moves towards left on the X axis, the number of node points in the mesh increases. In general the trends for both $ C_L$ and $ C_D$ show that as more node points are added with refinement, better approximation with respect to the experimental values is achieved. For $C_D$ the best approximation to experiments is given by the fine mesh which has the largest number of node points. Curiously, for $ C_L$, medium mesh gives a better approximation than the fine mesh and fine mesh over predicts $ C_L$. The adapted mesh which includes the same order of node points as of the coarse mesh, does a much better job of predicting both $ C_L$ and $ C_D$ than the coarse mesh. 

For 16\degrees, $ C_L$ shows the same behavior where the fine mesh over predicts $ C_L$ but with a closer approximation to the experimental value than the other meshes. The $ C_D$ curve shows a clear trend of moving towards the experimental value as the mesh is refined. The adapted results again show closer approximations to the experimental values than the coarse mesh. 

\subsubsection{Pressure coefficients}

\begin{figure}[h!]
\begin{center}
\subfigure[Slat element: 29\% span] {
	\includegraphics[width=5.7cm]{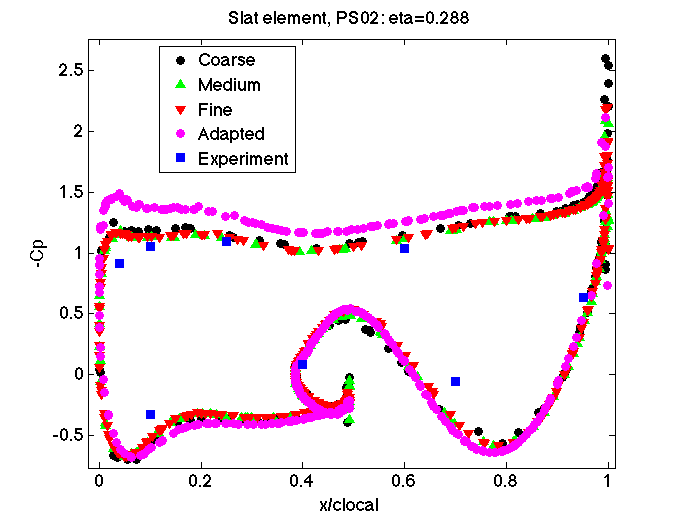}
	\label{f:DLRSlat17}
}
\hspace{-25pt}
\subfigure[Slat element: 68\% span] {
	\includegraphics[width=5.7cm]{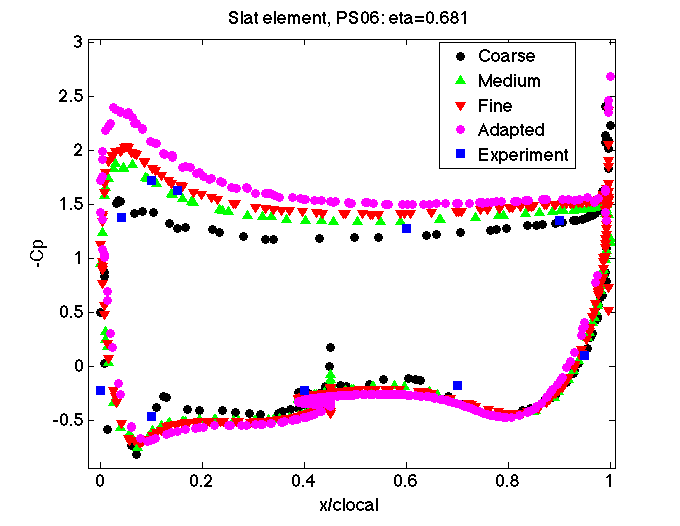}
	\label{f:DLRSlat50}
}
\hspace{-25pt}
\subfigure[Slat element: 89\% span]{
	\includegraphics[width=5.7cm]{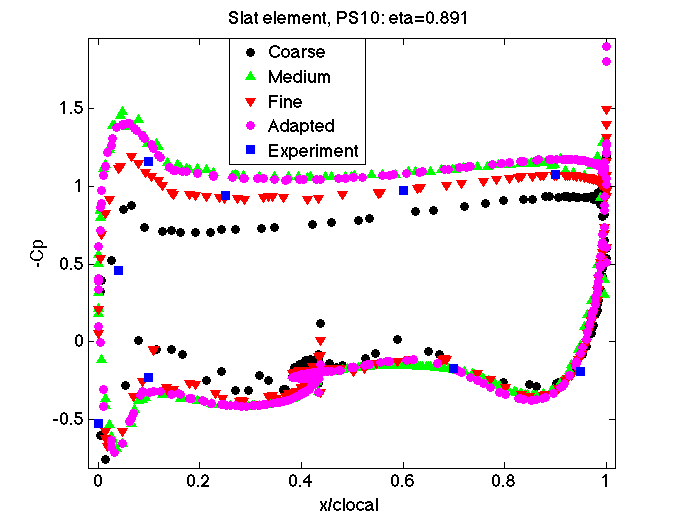}
	\label{f:DLRSlat98}
}
\hspace{-25pt}
\subfigure[Main wing: 29\% span] {
	\includegraphics[width=5.7cm]{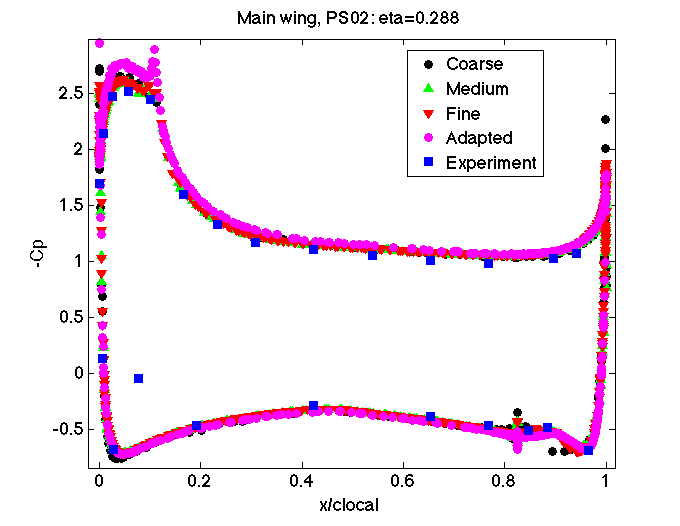}
	\label{f:DLRMain17}
}
\hspace{-25pt}
\subfigure[Main wing: 68\% span]{
	\includegraphics[width=5.7cm]{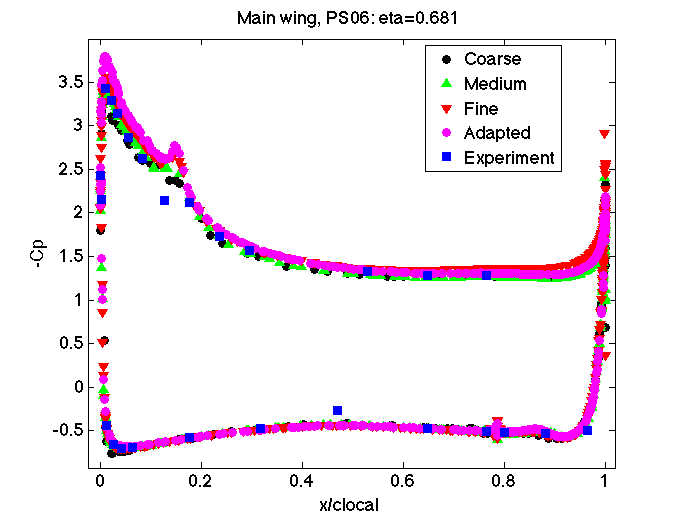}
	\label{f:DLRMain50}
}
\hspace{-25pt}
\subfigure[Main wing: 89\% span]{
	\includegraphics[width=5.7cm]{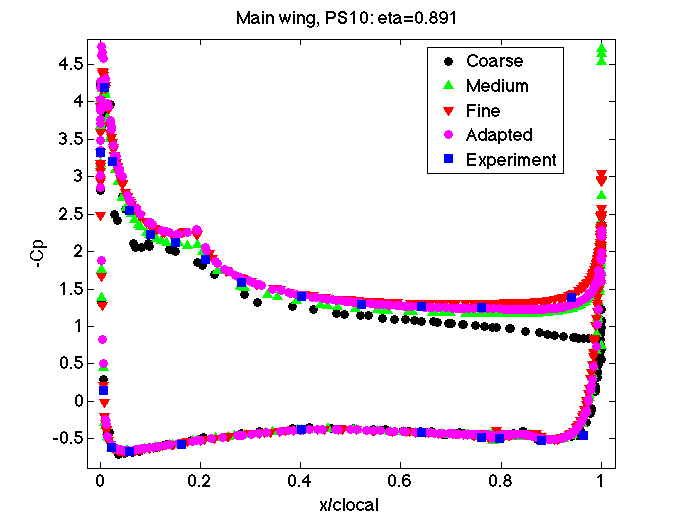}
	\label{f:DLRMain98}
}	
\hspace{-25pt}
\subfigure[Flap element: 29\% span] {
	\includegraphics[width=5.7cm]{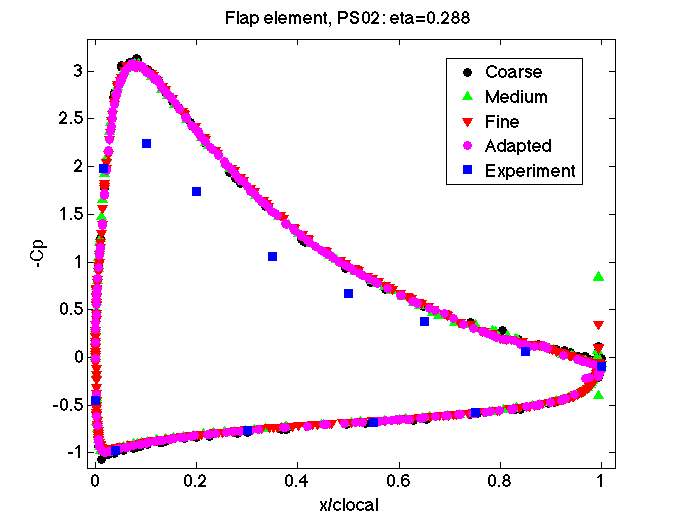}
	\label{f:DLRFlap17}
}
\hspace{-25pt}
\subfigure[Flap element: 68\% span]{
	\includegraphics[width=5.7cm]{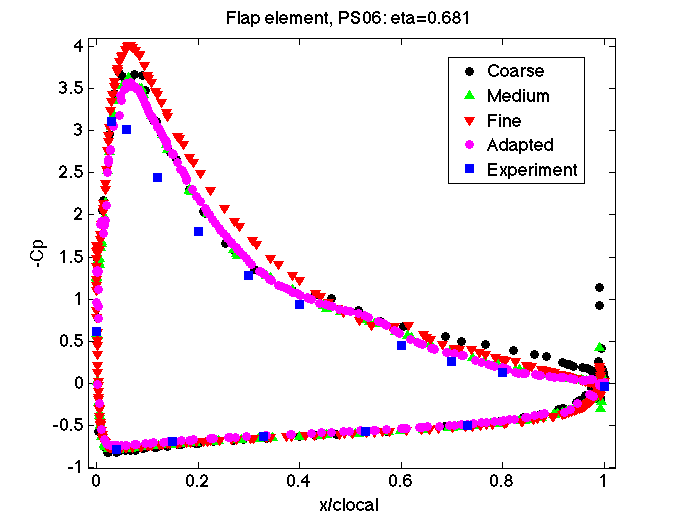}
	\label{f:DLRFlap50}
}
\hspace{-25pt}
\subfigure[Flap element: 89\% span]{
	\includegraphics[width=5.7cm]{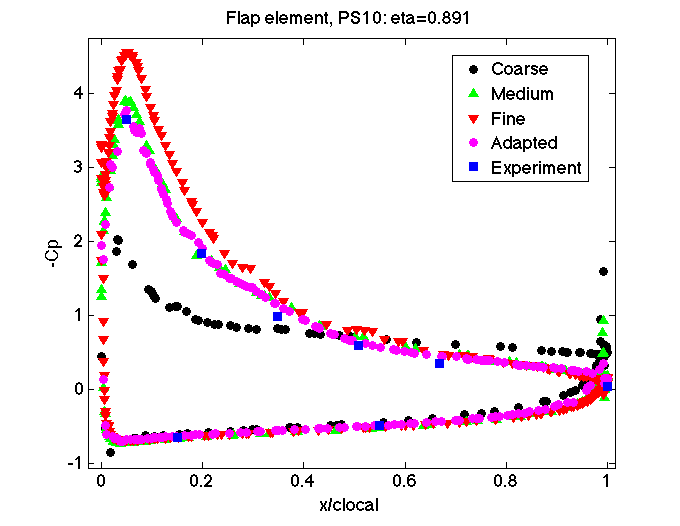}
	\label{f:DLRFlap98}
}
 \end{center}
\vspace{-15pt}
 \caption{Coefficient of pressure at 29\%, 68\%and 89\% span for Case 1, 7\degrees \hspace{0.1mm} AoA}
 \label{f:DLRCp1750All}
\end{figure}

Figure~\ref{f:DLRCp1750All} plots the coefficient of pressure at the 29\%, 68\% and 89\% spanwise sections on respectively the slat, the main wing and the flap. Note that $ -C_p$ is plotted on the y-axis. 

For the slat, the adapted mesh over predicts the suction pressure relative to other meshes but gives a better agreement than the coarse mesh on the pressure side. At 29\% span, all curves lie pretty much on top of each other except for the adapted results. At 68\% span, the coarse mesh under predicts the pressure on the suction side and the adapted mesh over predicts it. The medium and the fine mesh are in a good agreement with the experiments. However, the adapted mesh gives better agreement than the coarse mesh on the pressure surface. 

For the main wing, all meshes give good agreement with the experimental results for the 29\% and 68\% sections. The adapted mesh over predicts the suction peak by a small amount. A good thing to note is that all the meshes are able to capture the effect of separation on the pressure, near the trailing edge. 

All meshes give deviating results from the experiments for the flap 29\% section, but agree well with each other. This indicates that we have mesh convergence but the turbulence model is unable to capture the effect of separation near the root of the wing, leading to the conclusion that we have verified results for this particular turbulence model. For the 68\% section of the flap, the fine mesh overshoots the pressure peak on the suction side as compared to other meshes, but other meshes agree with each other. Near the trailing edge, the coarse mesh predicts more separation compared to the experiments and other meshes. 

At 89\% span, the coarse mesh shows degrading performance and does not give matching values to the experiments for all the three elements. For the slat, it under predicts the pressure on the suction surface for all of the section. Interestingly, the fine mesh gives a good agreement with the experimental data but the medium and the adapted mesh both over predict the pressure on the suction side by a small amount. On the pressure surface of the slat as well, adapted mesh performs better than the coarse mesh. 

For the 89\% section of the main wing section (Figure~\ref{f:DLRMain98}), the coarse mesh gives continuous increasing pressure (sustained adverse pressure gradient) on the suction side near the trailing edge indicating attached behavior which does not agree with the experiments. The medium, fine and the adapted mesh are all in good agreement with the experiments and capture the effect of separation near the trailing edge to a good degree. For the flap section, the coarse mesh does a poor job at capturing the suction pressure peak and gives a flatter curve which does not match the experiments. The medium and the adapted meshes give values that lie on top of each other and agree well with the experimental values. The fine mesh, however, overshoots the pressure peak slightly, indicating that these meshes bound the experimental values.e mesh was observed by other participants in the workshop, as well\cite{HiLiftPW2Summary}. This peculiar behavior of the fin. Near the trailing edge, except for the coarse mesh, other meshes show attached behavior like the experiments. Similar behavior of pressure coefficients was observed for 16\degrees \hspace{0.1mm} AoA. 

\subsubsection{Vorticity contours and velocity profiles}

\begin{figure}[h!]
\begin{center}
\subfigure[Coarse mesh]{
	\includegraphics[width=3.8cm]{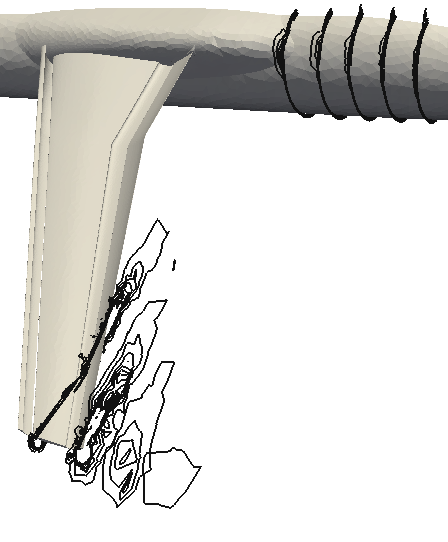}
	\label{f:DLRVortCoarse}
}
\subfigure[Medium mesh]{
	\includegraphics[width=3.8cm]{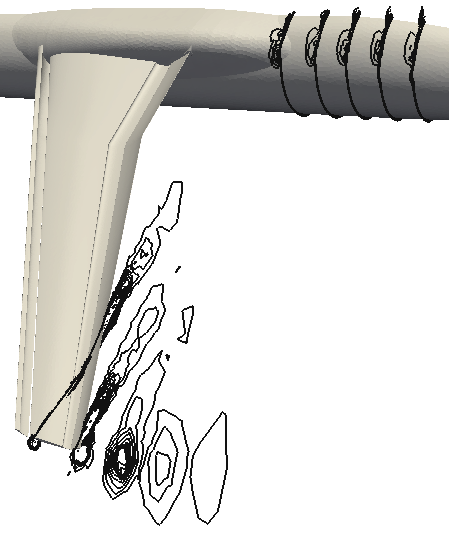}
	\label{f:DLRVortMedium}
}	
\subfigure[Fine mesh]{
	\includegraphics[width=3.8cm]{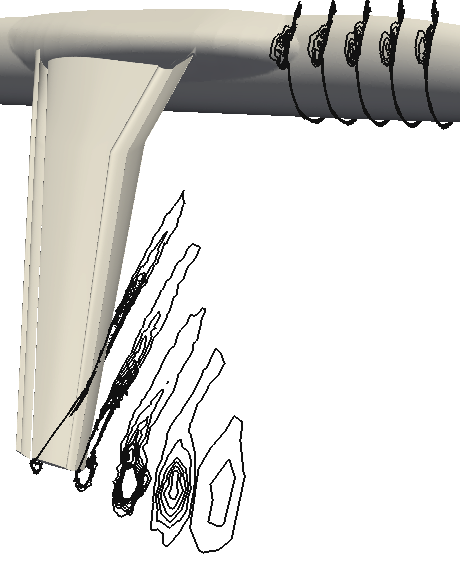}
	\label{f:DLRVortFine}
}
\subfigure[Adapted mesh]{
	\includegraphics[width=3.8cm]{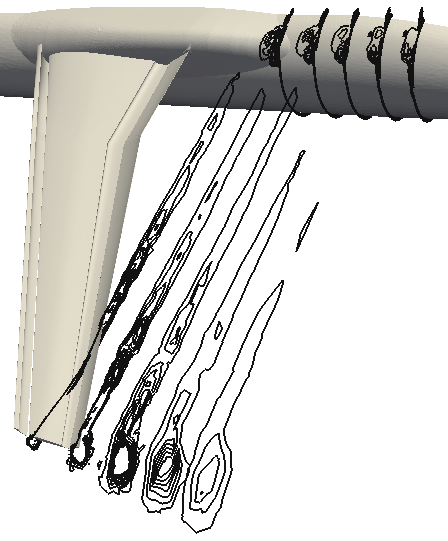}
	\label{f:DLRVortAdapt1}
}
\end{center}
\vspace{-15pt}
 \caption{Vorticity contours on planes with normals in X direction for Case 1, 7\degrees \hspace{0.1mm} AoA}
 \label{f:DLRVortCont}
\end{figure}

As seen from the $ C_p$ plots, one area where the adapted mesh does better than the coarse mesh is the tip region. We plot the vorticity contours on planes with X normals in Figure~\ref{f:DLRVortCont} for the coarse, medium, fine and the adapted meshes to show the development of the tip vortex. The adapted mesh undoubtedly gives better vorticity capturing in the tip region compared to the coarse and the medium meshes and on par with the fine mesh. Adaptive refinement in this region is high and the increased resolution leads to superior capturing of the flow features. The adapted mesh captures a lot of vortical activity in the wake of the wing as well which is missed by the other meshes. This comes naturally from the choice of error indicators as the wake and tip regions are refined heavily. Off body vortices along the lower side of the fuselage are captured to some extent by the medium mesh but are not clearly seen for the coarse mesh. The fine and the adapted meshes capture the off body features with comparable quality. Vorticity contours for 16\degrees \hspace{0.1mm} AoA show similar behavior. 

\begin{figure}[h!]
\begin{center}
	\includegraphics[width=12cm]{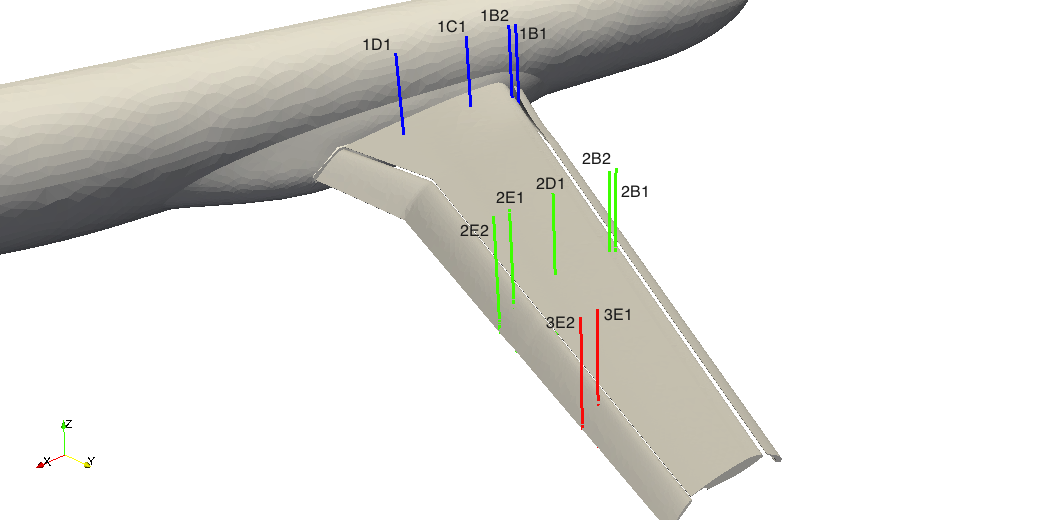}
	\caption{Locations and labeling of the Z constant lines along which the velocity data is extracted}
	\label{f:VelPlanes}
\end{center}
\end{figure}

\begin{figure}[h!]
\begin{center}
\subfigure[1B1]{
	\includegraphics[width=2.7cm]{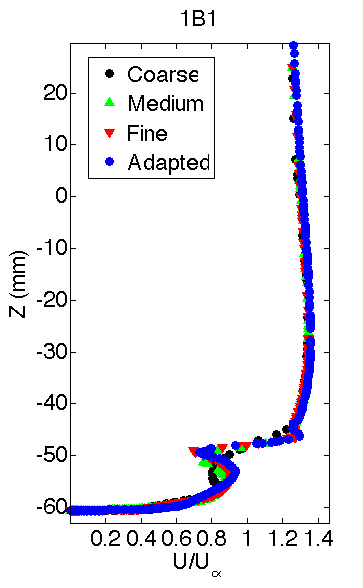}
	\label{}
}
\hspace{-15pt}
\subfigure[1B2]{
	\includegraphics[width=2.7cm]{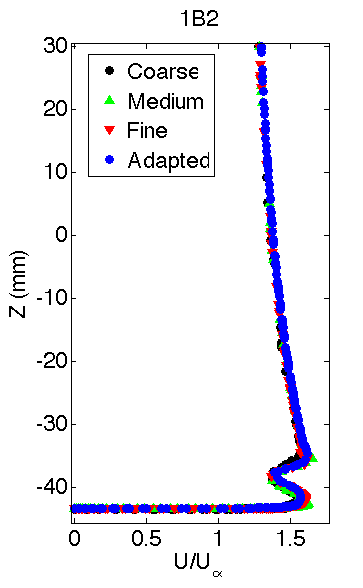}
	\label{}
}
\hspace{-15pt}
\subfigure[1D1]{
	\includegraphics[width=2.7cm]{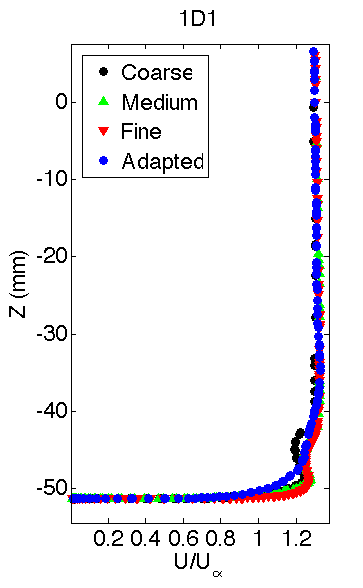}
	\label{}
}
\hspace{-15pt}
\subfigure[2B2]{
	\includegraphics[width=2.7cm]{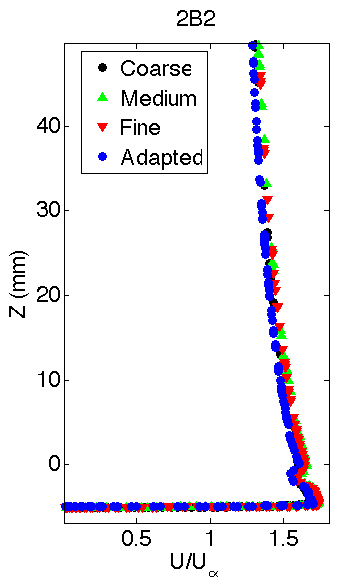}
	\label{}
}
\hspace{-15pt}
\subfigure[2E1]{
	\includegraphics[width=2.7cm]{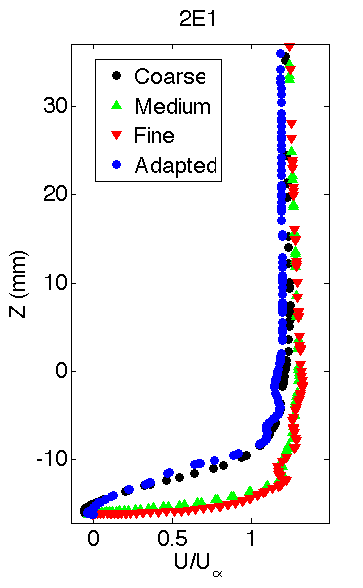}
	\label{}
}
\hspace{-15pt}
\subfigure[3E2]{
	\includegraphics[width=2.7cm]{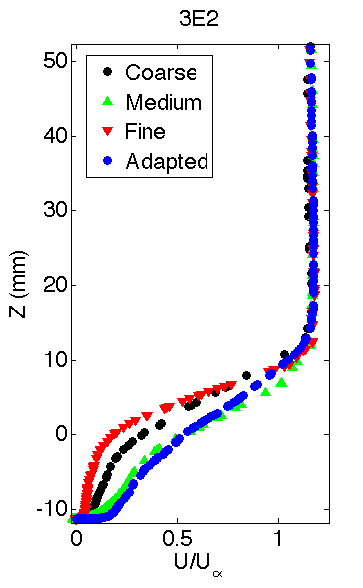}
	\label{}
}
\end{center}
\vspace{-15pt}
\caption{Normalized U Velocity plots for lines extracted parallel to the Z axis for Case 1, 7\degrees \hspace{0.1mm} AoA}
\label{f:7DegVel1}
\end{figure}

Figure~\ref{f:7DegVel1} plots the normalized U velocity profiles at various stations along the wing, parallel to the Z axis. The various lines with constant Z along which the data are extracted are shown in Figure~\ref{f:VelPlanes}. Particularly, the B stations are immediately in the slat wake, C and D stations are on the main wing, and E stations are on the flap, in the wake of the main wing. 

The simulations are able to capture the wake of the slat for the 1B1 and 1B2 stations which is marked by a concave bump in the velocity profile indicating a dip in the velocity. The 2B2 station also shows a slight wake effect. Overall there is a good agreement between profiles predicted by different meshes except for the stations in the wake of the main wing (E stations). The fine mesh shows deviation from other meshes on these stations.

\subsection{Reynolds number study (Case 2)}

This study was split in 2 parts; The first one was dedicated to a low Reynolds number (1.35 million) and the second one to a high Reynolds number (15.1 million). Several angles of attack were considered for both cases. One single mesh was used for the entire analysis of this case which was of comparable grid size to the medium mesh from the previous analysis presented in the previous section. Other details of the mesh have already been discussed in Section \ref{s:meshgen}.

\subsubsection{Low Reynolds number}

Case 2a was a study for a low Reynolds number of 1.35 million. This case was simulated for 8 different angles of attack: 0\degrees, 7\degrees, 12\degrees, 16\degrees, 18.5\degrees, 19\degrees, 20\degrees and 21\degrees. The calculations were made at a constant time step of 1e-4 $ s$ leading to unsteady RANS (URANS) simulations. Time averaged quantities were used to plot coefficient of pressure and lift and drag properties. 
 
\begin{figure}[h!]
\begin{center}
\subfigure[$ \mathrm{C_L \, vs. \: \alpha}$]{
	\includegraphics[width=5.6cm]{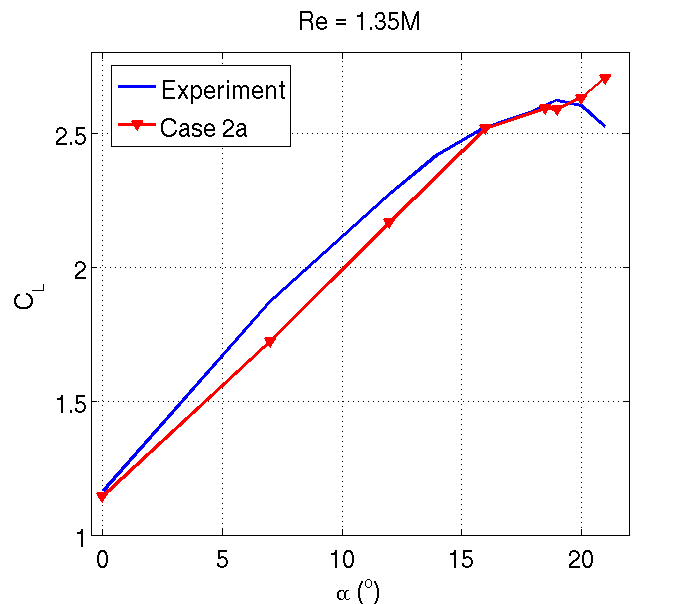}
	\label{f:CLalpha_2a}
}
\hspace{-25pt}
\subfigure[$ \mathrm{C_D \, vs. \: \alpha}$]{
	\includegraphics[width=5.6cm]{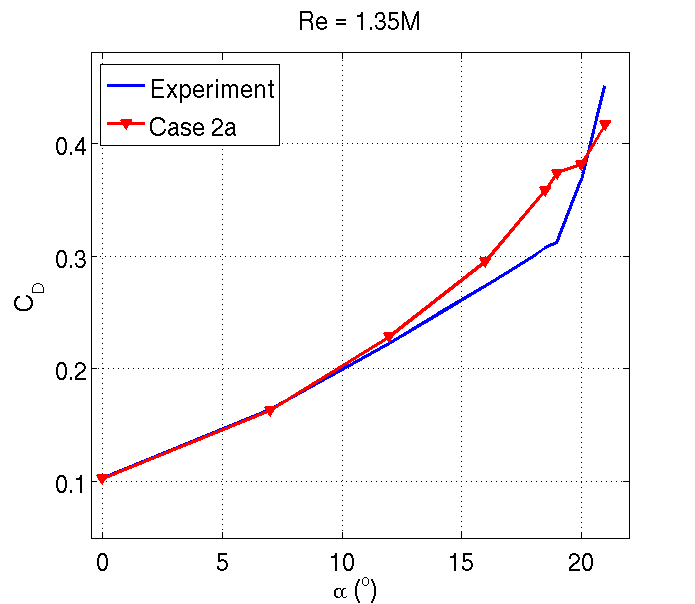}
	\label{f:CDalpha_2a}n
}
\hspace{-25pt}
\subfigure[$ \mathrm{C_D \, vs. \: C_L}$]{
	\includegraphics[width=5.6cm]{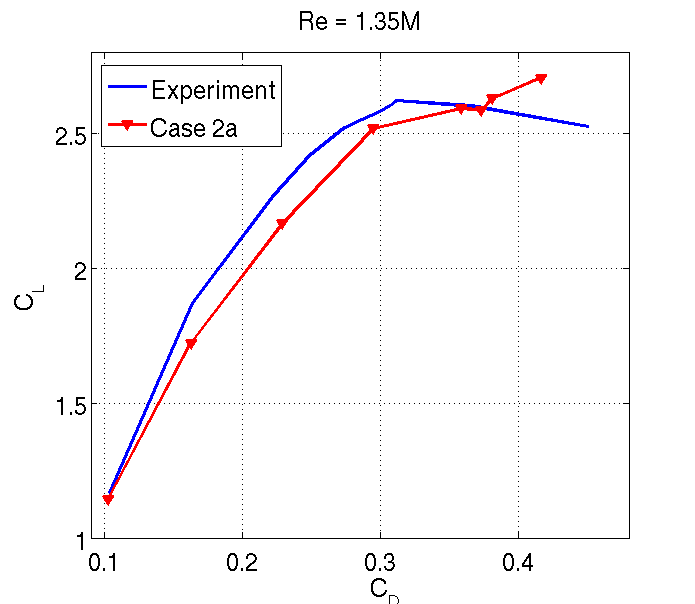}
	\label{f:CLCD_2a}
}
\end{center}
\vspace{-15pt}
\caption{Lift and drag plots for 1.35 million Reynolds number (Case 2a)}
\label{f:LDRe135}
\end{figure}

Figure~\ref{f:LDRe135} plots the lift and drag curves for Case 2a. Figure~\ref{f:CLalpha_2a} plots the coefficient of lift against the angle of attack. Simulation results are in a close proximity to the experiments till stall is reached. At 21\degrees \hspace{0.1mm} experiments predict beyond stall properties which are marked by a dip in the curve. However, our simulations are unable to capture this effect. Interestingly, the $C_L$ values near 7\degrees \hspace{0.1mm} and 12\degrees \hspace{0.1mm} differ from the experiments by a larger magnitude than near stall angle of attacks. Figure~\ref{f:CDalpha_2a} shows coefficient of drag vs. angle of attack curve. In general simulations predict higher drag than experiments especially for angle of attacks 16\degrees, 18.5\degrees \hspace{0.1mm} and 20\degrees. Figure~\ref{f:CLCD_2a} plots the drag polar for Case 2a. The simulations are able to capture the overall shape of the curve except for beyond stall, showing higher $ C_L$ than for the experiments. 

\begin{figure}[h!]
\begin{center}
\subfigure[1B1]{
	\includegraphics[width=2.7cm]{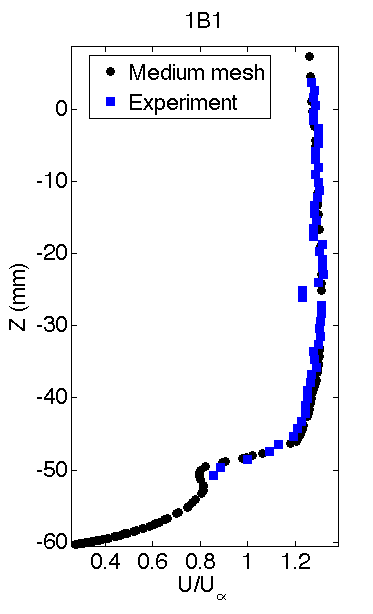}
	\label{}
}
\hspace{-15pt}
\subfigure[1B2]{
	\includegraphics[width=2.7cm]{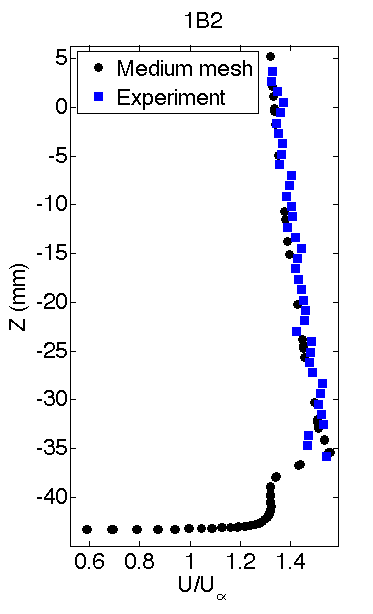}
	\label{}
}
\hspace{-15pt}
\subfigure[1D1]{
	\includegraphics[width=2.7cm]{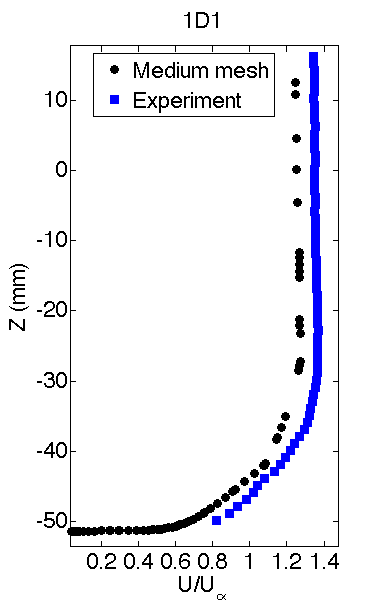}
	\label{}
}
\hspace{-15pt}
\subfigure[2E1]{
	\includegraphics[width=2.7cm]{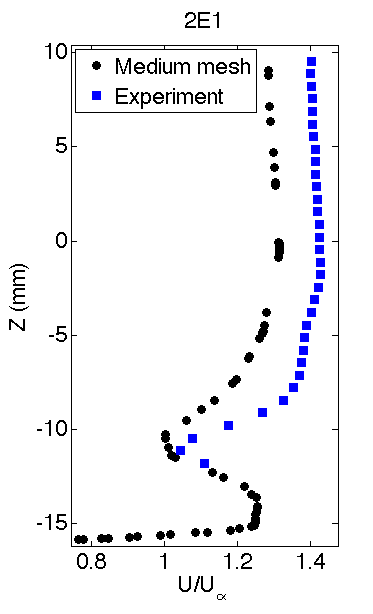}
	\label{}
}
\hspace{-15pt}
\subfigure[2E2]{
	\includegraphics[width=2.7cm]{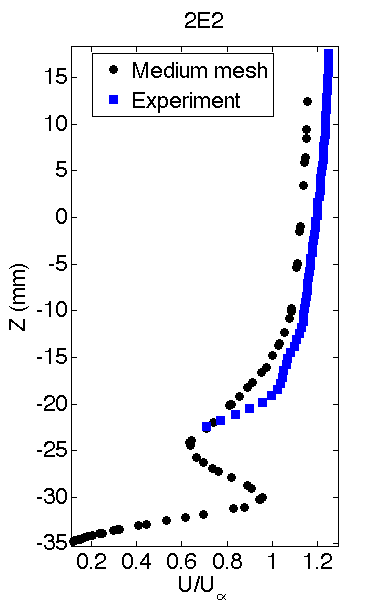}
	\label{}
}
\hspace{-15pt}
\subfigure[3E2]{
	\includegraphics[width=2.7cm]{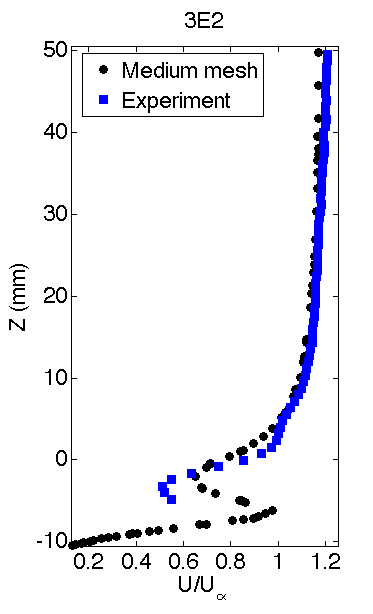}
	\label{}
}
\end{center}
\vspace{-15pt}
\caption{Normalized U Velocity plots for lines extracted parallel to the Z axis for Case 2a, 7\degrees \hspace{0.1mm} AoA}
\label{f:7DegVel2a}
\end{figure}

\begin{figure}[h!]
\begin{center}
\subfigure[1B1]{
	\includegraphics[width=2.7cm]{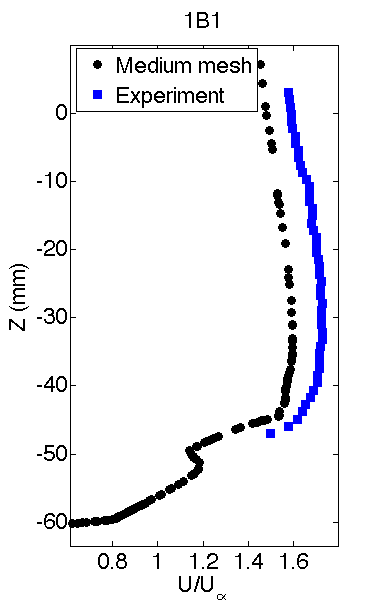}
	\label{}
}
\hspace{-15pt}
\subfigure[1B2]{
	\includegraphics[width=2.7cm]{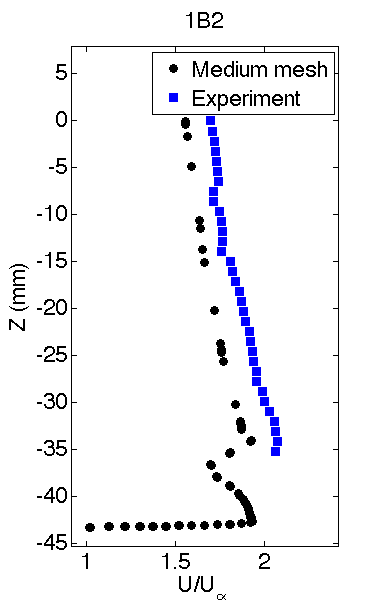}
	\label{}
}
\hspace{-15pt}
\subfigure[1D1]{
	\includegraphics[width=2.7cm]{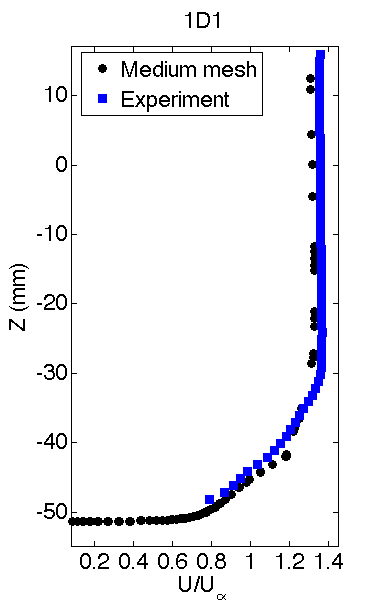}
	\label{}
}
\hspace{-15pt}
\subfigure[2B2]{
	\includegraphics[width=2.7cm]{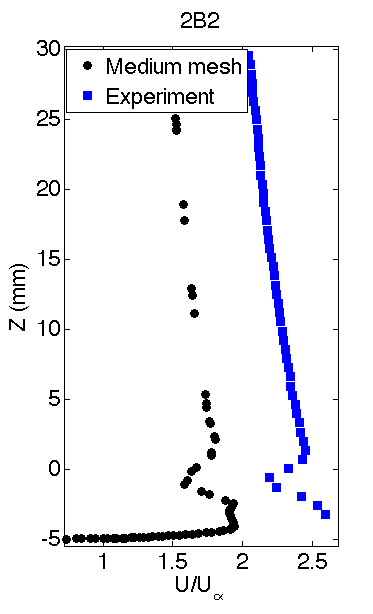}
	\label{}
}
\hspace{-15pt}
\subfigure[2D1]{
	\includegraphics[width=2.7cm]{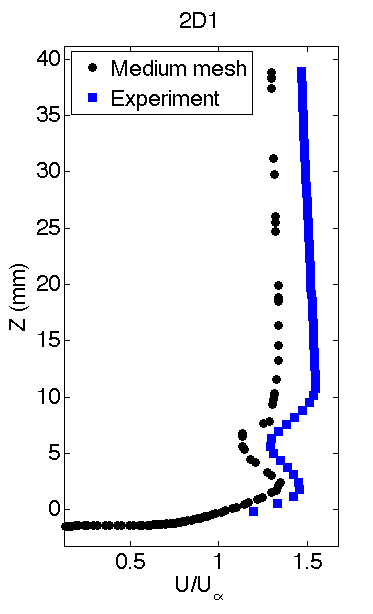}
	\label{}
}
\hspace{-15pt}
\subfigure[3E2]{
	\includegraphics[width=2.7cm]{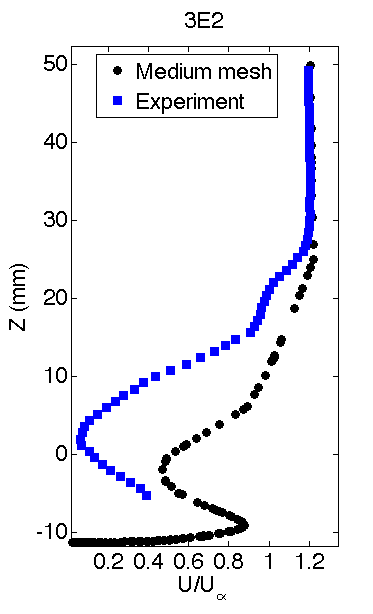}
	\label{}
}
\end{center}
\vspace{-15pt}
\caption{Normalized U Velocity plots for lines extracted parallel to the Z axis for Case 2a,  18.5\degrees \hspace{0.1mm} AoA}
\label{f:185DegVel2a}
\end{figure}

Normalized U velocity is plotted in Figures~\ref{f:7DegVel2a} and~\ref{f:185DegVel2a} for various Z constant lines shown in Figure~\ref{f:VelPlanes}. For 7\degrees, there is a fairly good agreement with the experimental results. The velocity profiles show slight wake like structures for the slat wake B sections. The profiles in the wake of the main wing show a more pronounced wake which is well captured by the simulations and agree well with the experimental results. For 18.5\degrees \hspace{0.1mm} there is more deviation between the experimental values and velocity profiles from our simulations. The agreement is still good for the section near the inboard region, but for the E section the agreement becomes worse, especially at the outboard sections. However, the simulations are able to capture the small wakes behind the slats to a good degree. At section 3E2, simulations predict a smaller wake than the experiments. 

\begin{figure}[h!]
\begin{center}
\subfigure[Surface LIC: 7\degrees]{
	\includegraphics[width=7.5cm]{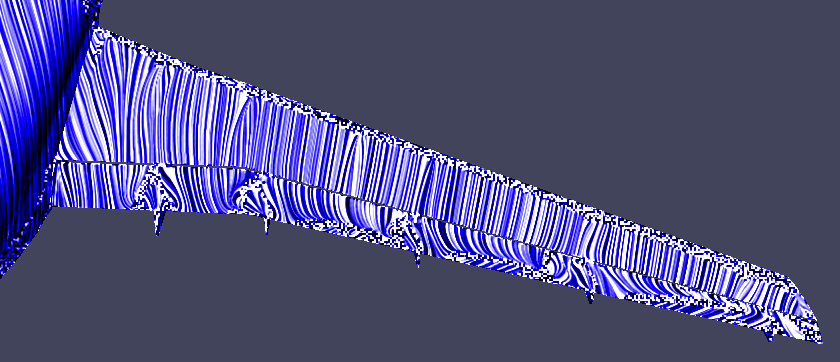}
	\label{f:7DegLIC}
}
\subfigure[Oil flow: 7\degrees]{
	\includegraphics[width=7.5cm]{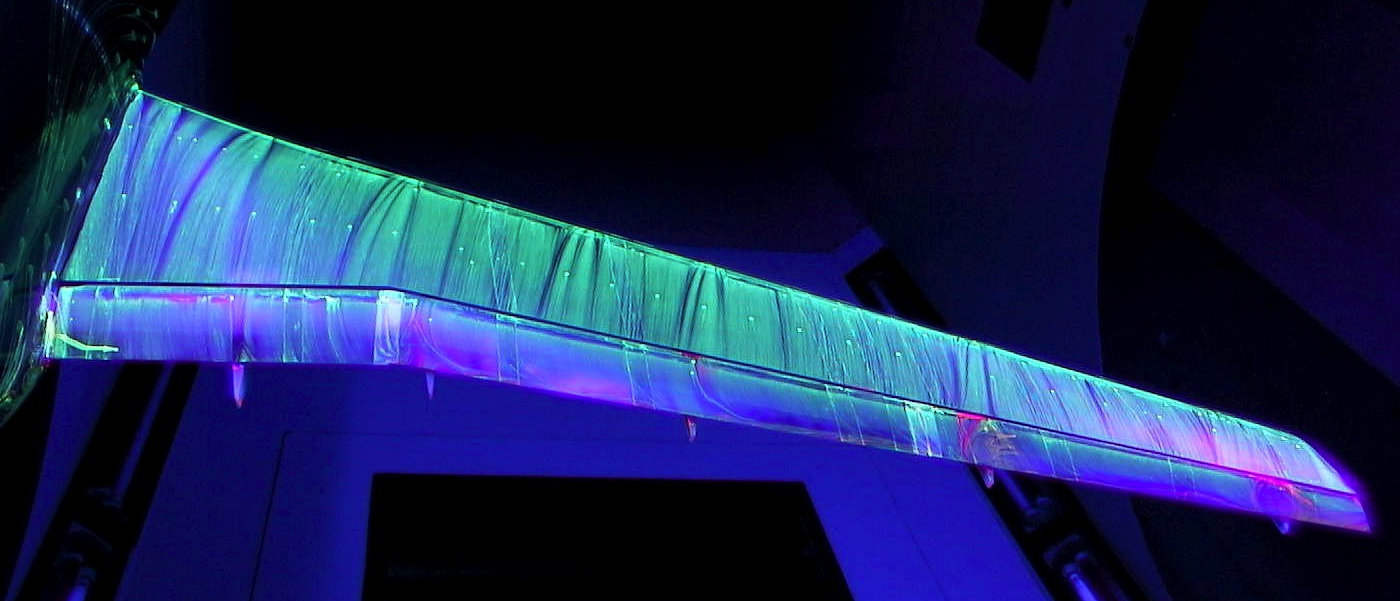}
	\label{f:7DegOil}
}
\subfigure[Surface LIC: 18.5\degrees]{
	\includegraphics[width=7.5cm]{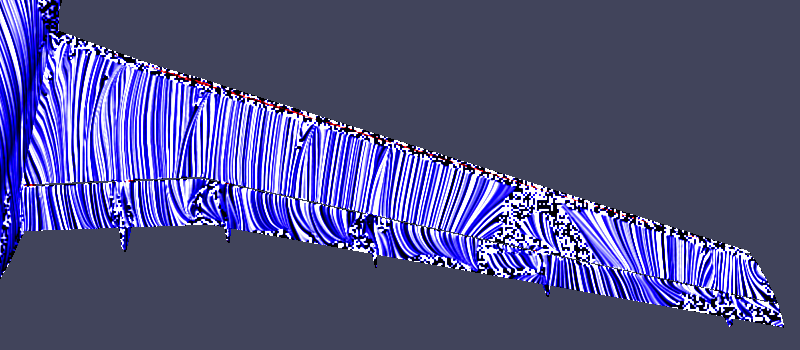}
	\label{f:185LIC}
}
\subfigure[Oil flow: 18.5\degrees]{
	\includegraphics[width=7.5cm]{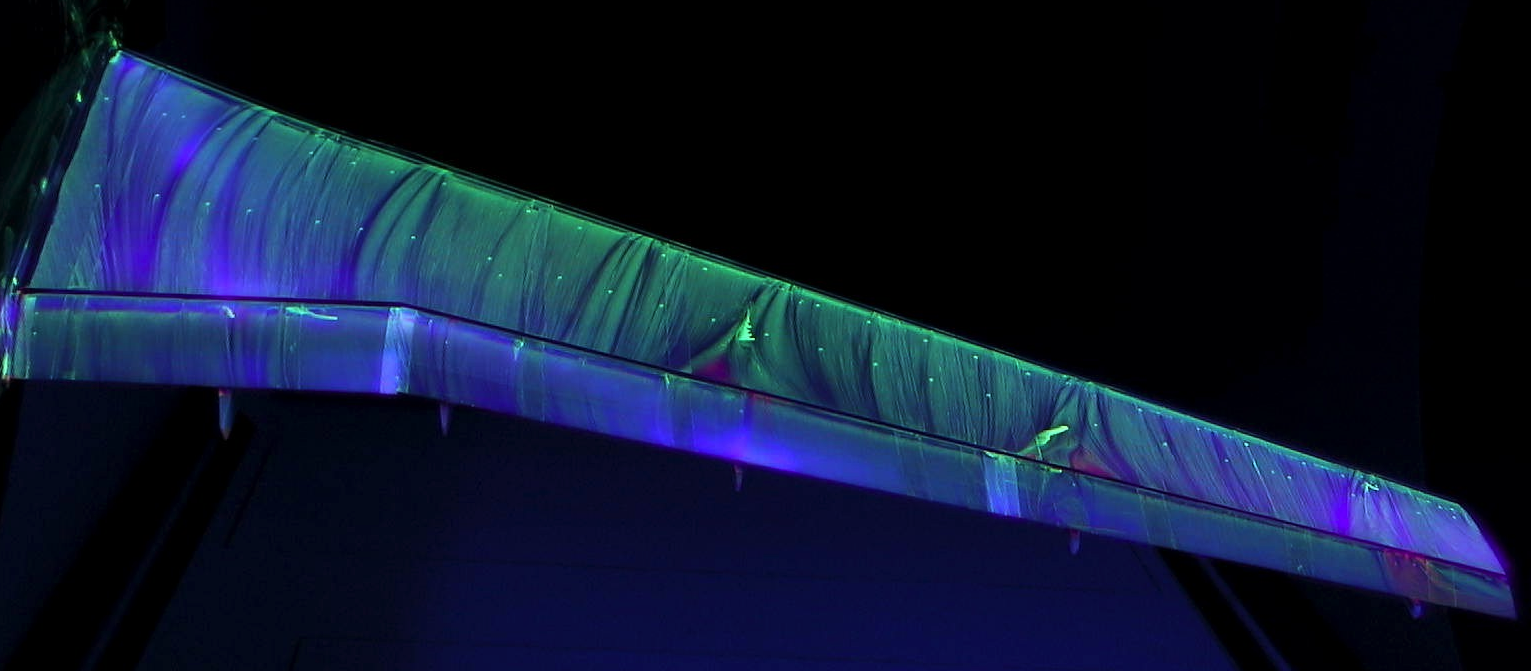}
	\label{f:185Oil}
}
\subfigure[Surface LIC: 21\degrees]{
	\includegraphics[width=7.4cm]{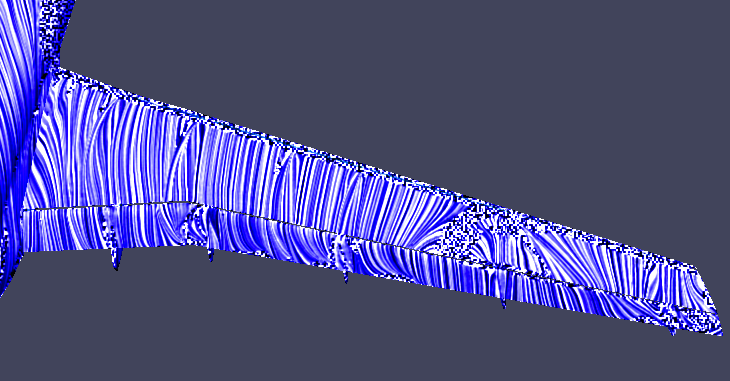}
	\label{f:21LIC}
}
\subfigure[Oil flow: 21\degrees]{
	\includegraphics[width=7.5cm]{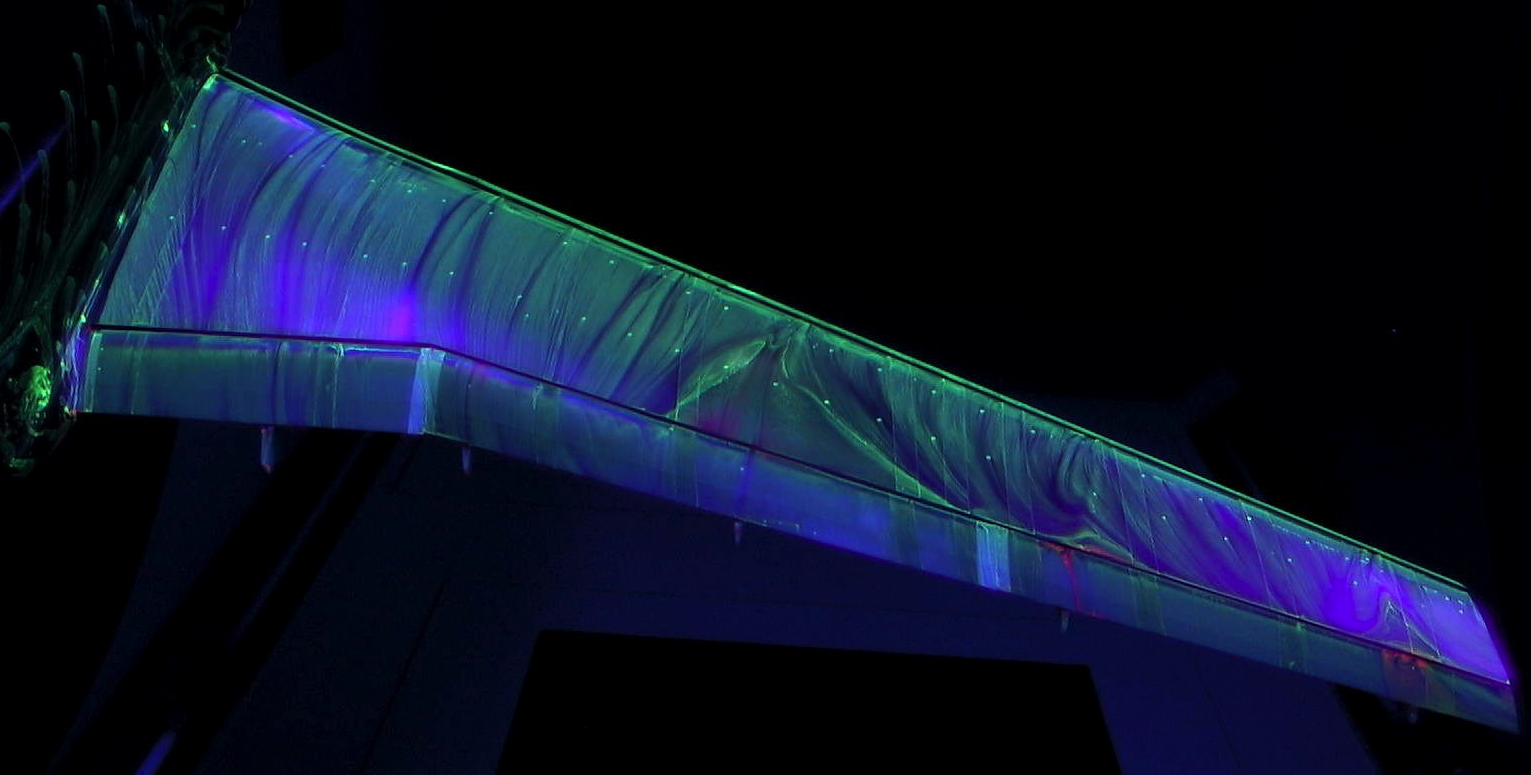}
	\label{f:21Oil}
}

\end{center}
\vspace{-15pt}
\caption{Lift and drag plots for 15.1 million Reynolds number (Case 2b)}
\label{f:OilFlow}
\end{figure}

Figure~\ref{f:OilFlow} shows comparison of surface LIC plots obtained with wall shear stress from the simulations with oil flow images from experiments.  At 7\degrees \hspace{0.1mm} AoA the flow relatively stays attached on most of the part of the main wing. This behavior is correctly captured by the simulations. The disturbances caused to the streamlines behind the slat tracks can be seen as well. At 18.5\degrees, oil flow image show 2 big disturbances which look like small separation zones behind a midboard and an outboard slat track. The simulations are not able to capture the mid board behavior but capture the outboard zone. This disturbance further develops into a bigger separation zone as the angle of attack is increased to 21\degrees. The oil flow image at 21\degrees \hspace{0.1mm} shows a big separation zone near midboard but most of the participants from the workshop failed to capture this behavior and instead predicted one near outboard like our results show. 

\subsubsection{High Reynolds number}

Case 2b is a high Reynolds number study for Re = 15.1 million. Similar to Case 2a, 8 different angles of attack are considered: 0\degrees, 7\degrees, 12\degrees, 16\degrees, 18.5\degrees, 20\degrees, 21\degrees, 22.4\degrees. The flow simulations were run at a time step of 1e-4 sec. 

\begin{figure}[h!]
\begin{center}
\subfigure[$ \mathrm{C_L \, vs. \: \alpha}$]{
	\includegraphics[width=5.5cm]{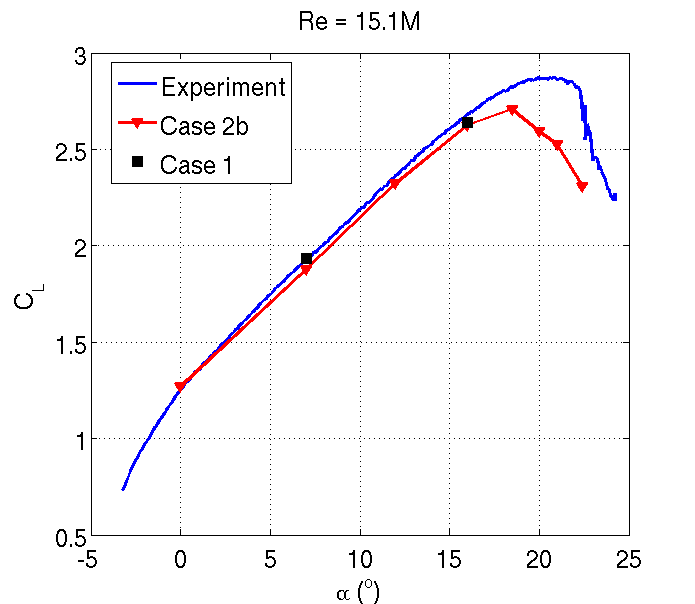}
	\label{f:CLalpha_2b}
}
\hspace{-25pt}
\subfigure[$ \mathrm{C_D \, vs. \: \alpha}$]{
	\includegraphics[width=5.5cm]{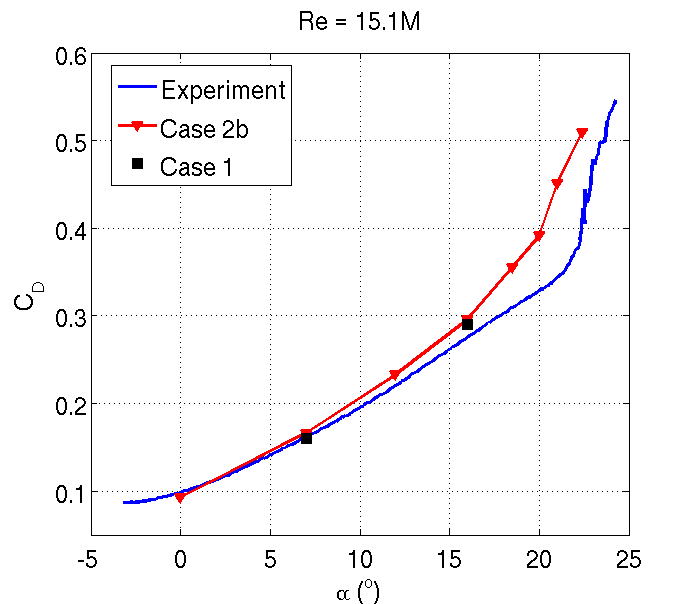}
	\label{f:CDalpha_2b}
}
\hspace{-25pt}
\subfigure[$ \mathrm{C_L\, vs. \: C_D}$]{
	\includegraphics[width=5.5cm]{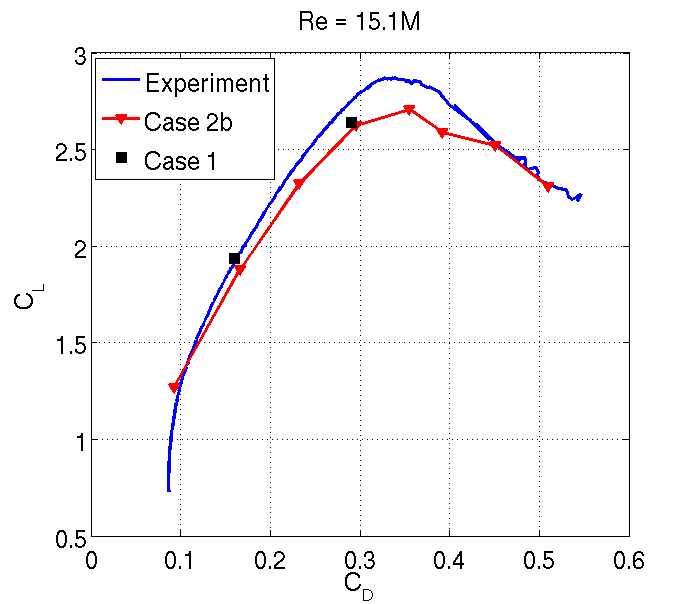}
	\label{f:CLCD_2b}
}
\end{center}
\vspace{-15pt}
\caption{Lift and drag plots for 15.1 million Reynolds number (Case 2b)}
\label{f:LDRe15}
\end{figure}

The trends for the lift and drag curves for Case 2b are similar to those from Case 2a. Figure~\ref{f:CLalpha_2b} plots $ C_L$ against $ \alpha$. For comparison we also plot values for Case 1 for 7 and 16\degrees. Our results follow the lift curve very well till about $ \alpha$ of 18.5. The predictions differ after that from the experimental values but the shape of the curve is still captured nicely, and the effect of stall is captured by a dip in the $ C_L$ values. Figure~\ref{f:CDalpha_2b} plots $ C_D$ vs. $ \alpha$. Overall our results predict higher drag especially at higher angles of attack. However, the shape of the curve is again captured to a good degree. Figure~\ref{f:CLCD_2b} shows the drag polar whose behavior is again captured to a good degree by the simulations. 

\subsubsection{Effect of the time step size $ \Delta t$}

\begin{figure}[h!]
\begin{center}
\subfigure[$ \mathrm{C_L \, vs. \: \alpha}$]{
	\includegraphics[width=5.7cm]{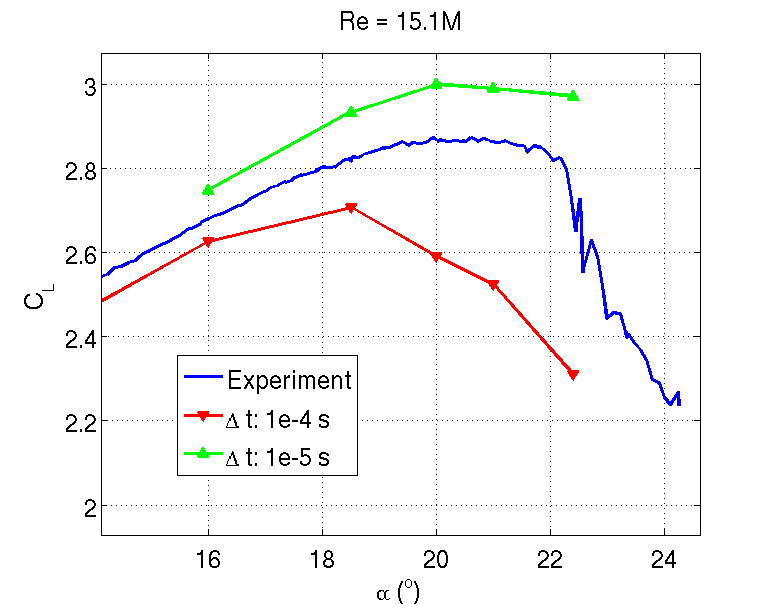}
	\label{f:CLalpha_2btime}
}
\hspace{-25pt}
\subfigure[$ \mathrm{C_D \, vs. \: \alpha}$]{
	\includegraphics[width=5.7cm]{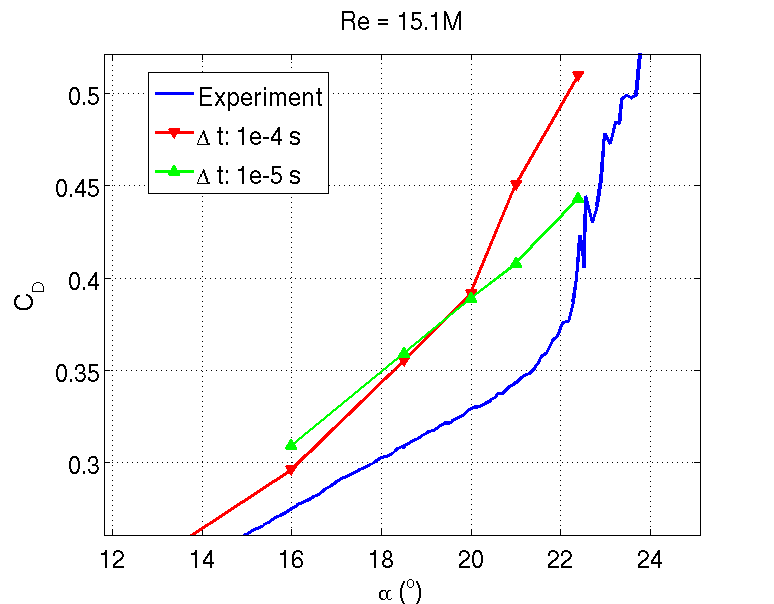}
	\label{f:CDalpha_2btime}
}
\hspace{-25pt}
\subfigure[$ \mathrm{C_L\, vs. \: C_D}$]{
	\includegraphics[width=5.7cm]{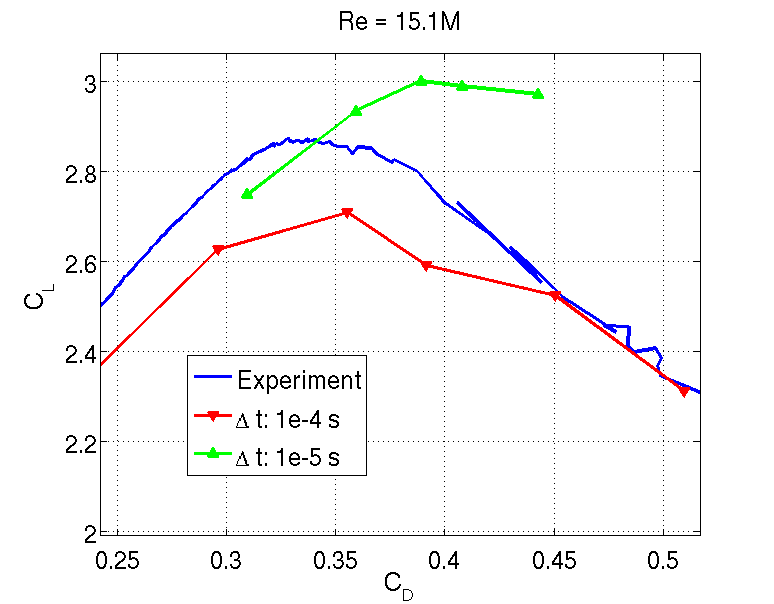}
	\label{f:CLCD_2btime}
}
\end{center}
\vspace{-15pt}
\caption{Lift and drag plots for a 15.1 million Reynolds number showing influence of $\Delta t$}
\label{f:LDRe15Time}
\end{figure}

All the simulations were run in a time accurate unsteady manner. This tends to cause some time step dependence of the solution, especially at higher angles of attack where most of the flow structures are highly unsteady. For Case 2b, since the values of $ C_L$ did not agree with the experiments for higher angles of attack, we ran those cases at a lower time step of 1e-5 $ s$. This change in the value of the time step showed significant change in the lift curve but did not show a major difference in the drag curve. Figure~\ref{f:LDRe15Time} compares the experimental lift and drag coefficients with our numerical results obtained with a time step of 1e-4 $ s$ and 1e-5 $s $ for 18.5\degrees, 20\degrees \hspace{0.1mm} and 21\degrees AoA. The figures are zoomed in to focus on the high angles of attack. 

The green curve follows the values obtained at a lower time step of 1e-5 $s$ as compared to the red curve obtained at 1e-4 $ s$. At 16\degrees, no significant difference in the results is noticeable. However, as we move towards stall, the lower time step predicts higher $ C_L$ than the higher time step. The two curves in a way bound the experimental curve. Interestingly, not much of difference is seen in the drag curve except at 21\degrees \hspace{0.1mm} and 22.4\degrees. At these angles of attack $ \Delta t = \text{1e-4} \; s$ predicts higher drag then $ \Delta t = \text{1e-5} \; s$. Naturally, the drag polar also shifts towards right above the experimental curve due to higher lifts at similar drag values. The drag polar for $ \Delta t = \text{1e-5} \; s$ is unable to predict the drop in the lift at higher drag values appropriate for 21\degrees \hspace{0.1mm} and 22.4\degrees, a behavior which is captured by the higher time step. This suggests highly sensitive nature of the flow to time stepping at higher angles of attack and requires further study with higher fidelity turbulence models than RANS or URANS like DES.

\subsubsection{Effects of slat tracks and flap fairings}

In this section we study the effects of slat tracks and flap fairings on properties like lift, drag and coefficient of pressure. Table~\ref{t:EffectCL} compares $ C_L$ and $ C_D$ for Case 1 and Case 2b with experiments for 7\degrees \hspace{0.1mm} angle of attack. It can be easily seen that Case 1 with no slat tracks and flap fairings predicts a higher $ C_L$ and a lower $ C_D$ than Case 2b. This is an expected result since slat tracks and flap fairings contribute towards friction drag. Similar trend is seen for 16\degrees \hspace{0.1mm} angle of attack. 

\begin{table}[h!]
\caption[]{Effect of slat tracks and flap fairings on $ \mathrm{C_L}$ and $ \mathrm{C_D}$: 7\degrees \hspace{0.1mm} AoA}
\newcolumntype{A}{>{\centering\arraybackslash}m{4 cm}}
\newcolumntype{B}{>{\centering\arraybackslash}m{3 cm}}
\newcolumntype{C}{>{\centering\arraybackslash}m{2 cm}}
\begin{tabular}{|C|B|C|}
\hline
 & $\mathrm{C_L}$ & $\mathrm{C_D} $ \\ 
\hline
Case 1 & 1.923 & 0.160\\
\hline
Case 2b &  1.876 & 0.166 \\
\hline
Experiment & 1.930 & 0.162 \\
\hline
\end{tabular}
\label{t:EffectCL}
\end{table}

\begin{figure}[h!]
\begin{center}
\subfigure[Slat element]{
	\includegraphics[width=5.7cm]{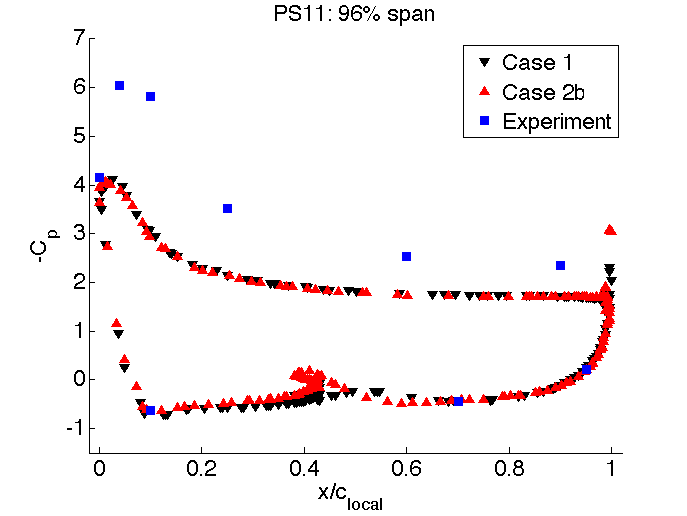}
	\label{f:EffectCpSlat}
}
\hspace{-25pt}
\subfigure[Main wing]{
	\includegraphics[width=5.7cm]{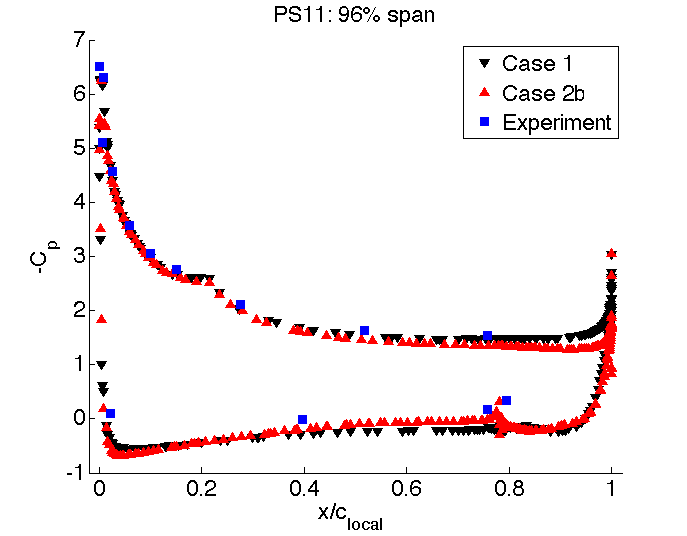}
	\label{f:EffectCpMain}
}
\hspace{-25pt}
\subfigure[Flap element]{
	\includegraphics[width=5.7cm]{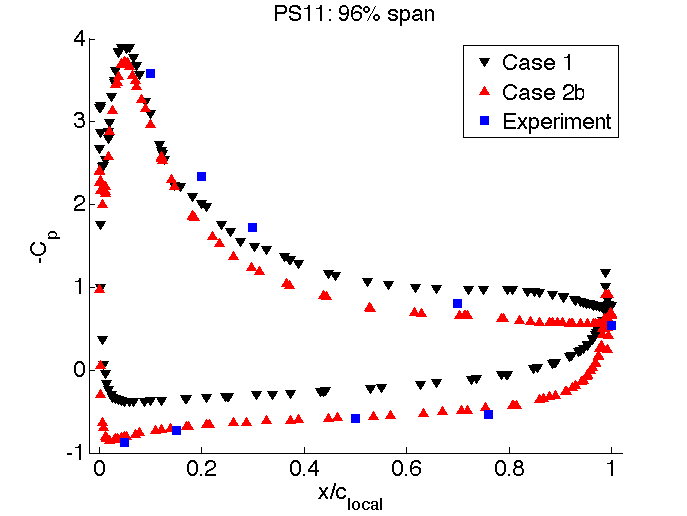}
	\label{f:EffectCpFlap}
}
\end{center}
\vspace{-15pt}
\caption{Comparison of $ C_p$ for Case 1 and Case 2b at 16\degrees \hspace{0.1mm} AoA on 96\% span section}
\label{f:EffectCp}
\end{figure}

Figure~\ref{f:EffectCp} shows coefficient of pressure plots at 96\% span section for 16\degrees \hspace{0.1mm} angle of attack. Expectedly, coefficient of pressure on the slat element was not significantly affected by slat tracks and flap fairings. On the main wing as well, not a lot of differences were seen between Case 1 and Case 2b. However, some differences were seen at 96\% span section especially on the pressure side of the wing which is very close to one of the flap fairings. In general, Case 2b under predicted the suction peak at most of the sections by a small amount. Larger differences between $ C_p$ values were seen on the flap sections for Case 1 and Case 2b. Again, the suction peak was under predicted on Case 2b. For 96\% span section, Case 1 did not agree with the experiments on the pressure side, where Case 2b was in a much better agreement with the experiments. 

\begin{figure}[h!]
\begin{center}
\subfigure[Case 1]{
	\includegraphics[width=7cm]{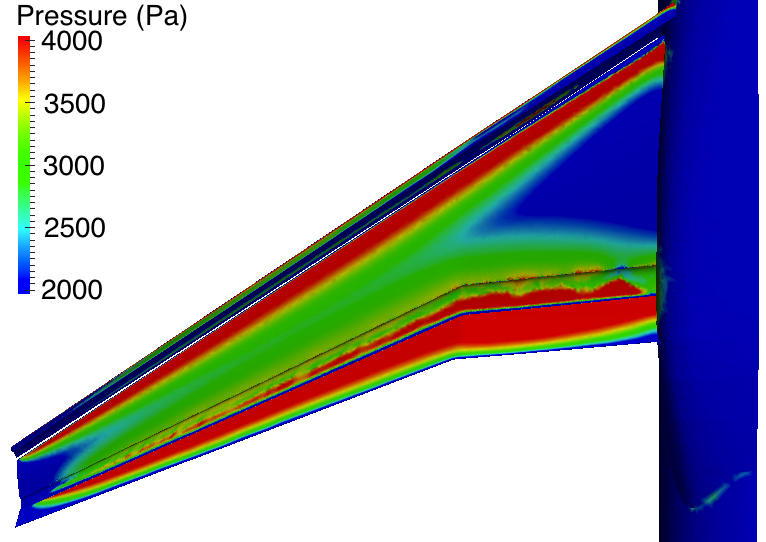}
	\label{f:EffectConfig2}
}
\subfigure[Case 2b]{
	\includegraphics[width=7cm]{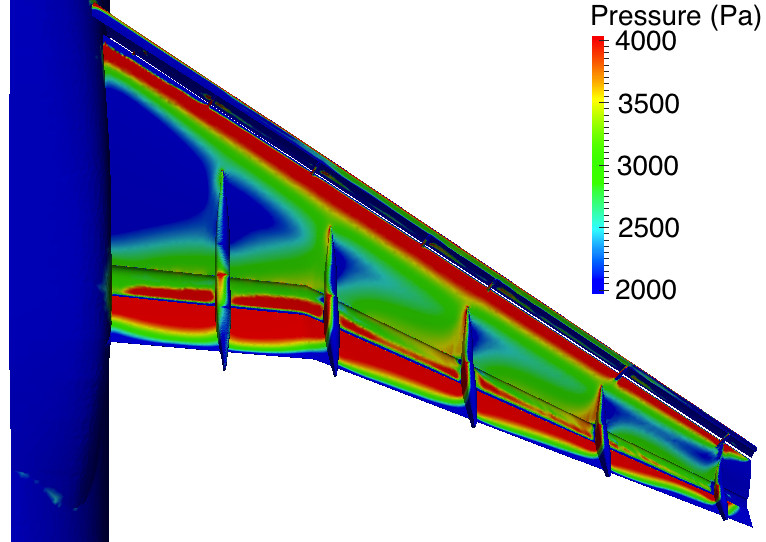}
	\label{f:EffectConfig4}
}
\end{center}
\vspace{-15pt}
\caption{Pressure distribution for Case 1 and Case 2b on the pressure side of the wing at 7\degrees \hspace{0.1mm} AoA}
\label{f:EffectPressure}
\end{figure}

The effect of the slat tracks and flap fairings on overall pressure distribution on the pressure side of the wing for 7\degrees \hspace{0.1mm} AoA is shown in Figure~\ref{f:EffectPressure}. The pressure field is affected especially near the flap fairings. Looking at the overall effects of slat tracks and flap fairings, one feels that they are an essential part of the configuration and should not be neglected for any full configuration study and comparison with experiments. 

\section{Conclusion}
\label{s:conclusions}

As a part of the $ 2^{\text{nd}}$ high lift prediction workshop, two studies related respectively to the grid convergence and the influence of the Reynolds number were completed. In house meshes were constructed following the meshing guidelines of the workshop. As expected, the grid convergence study showed a favorable trend towards grid convergence with increasing resolution. The coefficient of pressure on the wing surface and the velocity profiles were overall in a good agreement with the experiments. Anisotropic adaptive analysis showed a good potential of superior capturing of the flow structures at a reduced cost in terms of number of elements. The simulations were able to capture the right behavior of $ C_L$ and $ C_D$ with a fair comparison to the experiments. However, higher angles of attack showed a considerable dependence on the time step size. Including the slat tracks and flap track fairings resulted in increased drag and reduced lift and affected the coefficient of pressure particulary on the flap element. Because of the unsteady nature of the flow at high angles of attack, this case is a good candidate for detached eddy simulation and adaptivity.  
%

\section*{Acknowledgments}

This work is the results of several fruitful collaborations and we gratefully acknowledge Simmetrix Inc. for their meshing and geometric modeling libraries, Acusim Software Inc. (acquired by Altair Engineering) for their linear algebra solver library and Kitware for their visualization tools. This work utilized (i) the Mira BlueGene/Q supercomputer at ANL which is supported by the Department of Energy, (ii) the Janus supercomputer, which is supported by the National Science Foundation (award number CNS-0821794) and the University of Colorado Boulder.

\bibliography{bibiliography}
\bibliographystyle{aiaa}

\end{document}